\documentclass[aps,twocolumn,pra,tightenlines,amsmath,amssymb,floatfix,showpacs,superscriptaddress]{revtex4-1}
\usepackage[dvips]{graphicx}
\usepackage[english]{babel}
\usepackage{hyperref}
\usepackage{tikz}

\begin{document}

\title{Majorana zero modes in spintronics devices}

\author{Chien-Te Wu}
\affiliation{James Franck Institute, University of Chicago, Chicago, Illinois 60637, USA}
\affiliation{Department of Electrophysics, National Chiao Tung University, Hsinchu 30010, Taiwan, Republic of China}
\author{Brandon M. Anderson}
\affiliation{James Franck Institute, University of Chicago, Chicago, Illinois 60637, USA}
\author{Wei-Han Hsiao} 
\affiliation{James Franck Institute, University of Chicago, Chicago, Illinois 60637, USA}
\author{K. Levin}  
\affiliation{James Franck Institute, University of Chicago, Chicago, Illinois 60637, USA}

\begin{abstract}
We show that topological phases should be realizable in readily available and well studied heterostructures.
In particular we identify a new class of topological materials
which are well known in spintronics: helical ferromagnet-superconducting 
junctions.
We note that almost all previous work on topological heterostructures
has focused on creating Majorana modes at the proximity interface
in effectively two-dimensional or one-dimensional systems. 
The particular heterostructures we address exhibit finite range proximity
effects leading to nodal superconductors with Majorana
modes localized well away from this interface.
To show this, we implement a Bogoliubov-de Gennes (BdG) proximity numerical
scheme, which importantly,
involves
two finite dimensions in a three dimensional junction.
Incorporating this level of numerical complexity serves to
distinguish ours from alternative numerical BdG approaches which  
are limited by generally assuming translational invariance
or periodic boundary conditions along multiple directions.
With this access to the edges,
we are then able to illustrate in
a concrete fashion the wavefunctions of Majorana zero modes,
and, moreover, address finite
size effects. In the process we
establish consistency with a simple analytical
model. 
\end{abstract}


\maketitle

\section{Background}
The field of topological superconductivity has generated
exotic physics that realizes ideas from fields as diverse as
high energy~\cite{PhysRevD.13.3398},
atomic~\cite{1674-1056-24-5-050502,PhysRevA.93.063606,PhysRevLett.109.105303}
and condensed matter physics ~\cite{fu}.
Underlying this excitement has been
the lofty pursuit of novel phases of matter, as well as
implementing new
methods for quantum computing~\cite{sau}.
In making these superconductors experimentally there has
been a central focus on materials derived
from the proximity effect, where there is a higher level
of experimental control.
In such heterostructures, the central requirements of
spin-orbit coupling~\cite{PhysRevLett.100.096407, PhysRevLett.104.040502, PhysRevLett.104.067001, PhysRevLett.105.077001, PhysRevLett.105.177002},
as well as a Zeeman field,
and superconducting pairing can be
configured artificially.
For the most part these proximity-coupled exotic superconductors
involve topological-insulators~\cite{fu} or semiconductors
with strong spin-orbit scattering~\cite{sau}.
In these systems, the pairing is associated with
a two-dimensional (2D) $p_x \pm i p_y$ phase and
under ideal circumstances this
can lead to the possibilities of observing the
elusive Majorana modes.

In this paper, our goal is to arrive at Majorana surface states in a
different class of topological superconductors:
nodal superconductors based on 
helical ferromagnet (F)-superconductor (S) junctions. 
These systems are readily available and well studied in
the spintronics community~\cite{wu1,robinson,chiodi}.
Here
we characterize proximity-induced topological phases and
related edge states 
by numerically solving the finite size
Bogoliubov-de Gennes (BdG) equations
and providing consistency with simple analytic arguments.
The spin correlated and metallic nature of the ferromagnet results in 
pairing that penetrates significantly into the ferromagnetic region.
This is in contrast to the focus in past topological superconductivity 
literature~\cite{PhysRevLett.100.096407, PhysRevLett.104.040502, PhysRevLett.104.067001, PhysRevLett.105.077001, PhysRevLett.105.177002},
where two or one dimensional superconductivity is proximity induced
only at the interface region in insulators and semiconductors.

The resulting topological superconductivity exhibits a rich 
nodal structure~\cite{2015ChPhB..24e0502L, PhysRevLett.115.265304, PhysRevA.93.063606}
which, importantly, is associated with the existence of one or more flat bands 
reflecting zero-energy surface states.
We argue here these latter correspond to
Majorana modes, which are
found to be localized at the sample edge and outside the
transition interface between the
superconductor and normal phases.

Specifically we address
holmium/superconductor (F/S)
layered structures~\cite{wu1,robinson}
which exploit the intrinsic conical order of Ho. It should
be noted that there are similarities here to topological
order in artificially created one-~\cite{Yaz1,Yaz2,Yaz3,Yaz4}
and two-dimensional spin
configurations~\cite{Yaz5}. The F/S junctions
contain a host of
interesting superconducting properties~\cite{Volkov3}. Among these are:
(i)
The presence~\cite{halterman1,halterman2,Volkov1}
of anomalously long-range, equal-spin and $s$-wave
pair correlations in the ferromagnet. 
(ii) The oscillatory nature of the
Cooper pair amplitudes in the ferromagnetic region, which relates
to Larkin-Ovchinnikov-Fulde-Ferrell (LOFF) physics~\cite{demler,halterman3,Buzdin}.
(iii) The observation that these triplet Cooper pairs
can exist only if electrons are
paired odd in time (or frequency)~\cite{bergeret,Volkov2}.

To incorporate these more complicated features of the F/S proximity
structures into the present topological
study, we introduce a BdG analysis in which
there are two finite dimensions in a three dimensional system.
In this way
the numerics is more sophisticated than in
alternative analyses \cite{Lababidi,dasSarmanew} in the literature
which assume periodic boundary conditions along several dimensions.
As a consequence, we are able to not only demonstrate the existence of zero
energy flat bands but also plot the associated Majorana wavefunctions
which are localized at the edges. 
Importantly, these Majorana effects appear on the magnetic Ho side
of a proximity junction in which
there is no
attractive
interaction, and hence no true superconducting order parameter.
We find, additionally, that
ferromagnetic correlations can tunnel into the S side where
there is no magnetic order parameter.

\begin{figure}[h]
\includegraphics[width=3.3in]{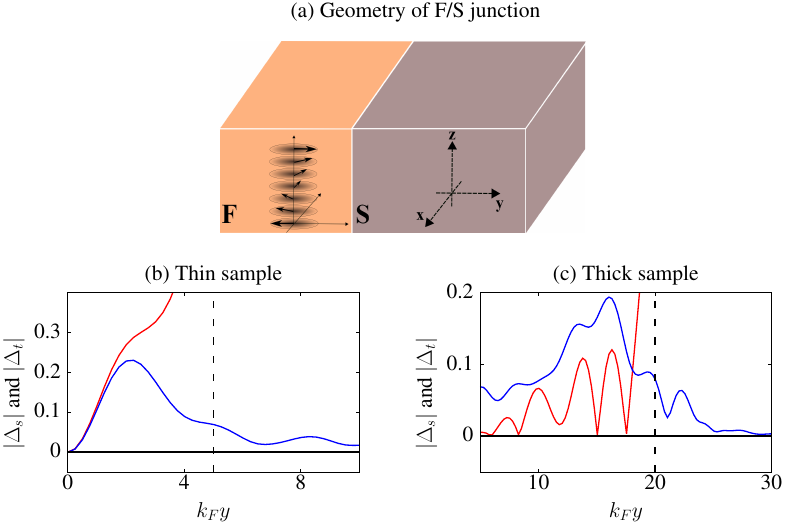}
\caption{
(a) The geometry of F/S bilayers: the interface lies on $x-z$ plane and has
finite extent along the $\hat{y}$ and $\hat{z}$ directions, and is infinite in the $\hat{x}$ direction.
(b) and (c) Proximity-induced singlet (red) and triplet (blue) correlations. 
These correlations penetrate significantly into the F (left of dashed lines) region due to the metallic structure.
Panel (c) considers a thicker F region ($y=20/k_F$) where LOFF oscillations of the singlet
component are observed, consistent with previous experiments. As shown in
Panel (b) for a more narrow F region 
considered in this paper, no oscillations are observed.
}
\label{fig:1}
\end{figure}

For definiteness we show the configuration of the F/S proximity junctions
in
the top panel of Fig.~\ref{fig:1}. Below 
we present
our numerical
results
(for two cases discussed below) for both singlet and triplet pair amplitudes. It 
can be seen 
that
the singlet pair amplitudes oscillate with a much shorter
decay length than their triplet counterparts.
Additionally the triplet correlations, which are spread throughout the ferromagnetic side, show
a small penetration into the superconducting region.
These long-range triplet correlations~\cite{linder}
were experimentally confirmed~\cite{robinson}, 
by introducing
Ho into Nb-based Josephson junctions and
observing a slow decay as a function of
ferromagnet thickness.

\section{Background theory}
Our system is described by a mean-field Hamiltonian
\begin{eqnarray} 
\label{eq:full}
\mathcal{H} &=& \int d^3 \mathbf{r} \,
 \psi_{\sigma}^{\dagger}(\mathbf{r}) \left( H_{\rm sp} + H_{\rm FM} \right)_{\sigma\sigma^\prime} \psi_{\sigma^\prime}(\mathbf{r}) \nonumber \\
 & & + \Delta(\mathbf{r})  \psi_{\uparrow}^{\dagger}(\mathbf{r}) \psi_{\downarrow}^{\dagger}(\mathbf{r}) +\textrm{h.c.},
\end{eqnarray}
involving fermions created (annihilated) by operators $\psi_{\sigma}^\dagger(\mathbf{r})$ ($\psi_{\sigma}(\mathbf{r})$) with spin $\sigma=\uparrow,\downarrow$. The single particle Hamiltonian $H_{\rm sp} = \left(-\nabla^2 / 2m - E_F \right)\sigma^0$ describes free fermions of mass $m$, and Fermi energy $E_F$; the identity operator $\sigma^0$ acts in spin space. Throughout we set $\hbar=k_B=1$.

The helical ferromagnet introduces
a coupling term $H_{\rm FM} = \mathbf{h}({\bf r}) \cdot \boldsymbol\sigma$, where $\boldsymbol\sigma = (\sigma^x, \sigma^y, \sigma^z)$ is a vector of Pauli matrices operating in spin space. 
The conical order of the exchange field ${\mathbf h}$, assumed to reside only in the ferromagnet, is written as
\begin{equation}
\label{exchange}
\mathbf{h}=h_0\left\{\cos\alpha\mathbf{\hat z}+\sin\alpha\left[ \sin\left(\frac
{\beta z}{a}\right)\mathbf{\hat x}+\cos\left(\frac{\beta z}{a}\right)\mathbf
{\hat y}\right]\right\},
\end{equation}
where $h_0$ is the internal field
strength of the ferromagnet. Here the helical ferromagnet parameters are set by a lattice constant along the $c$-axis of $a$, $\alpha \in [0,\pi]$ is the opening angle and $\beta$ sets the periodicity of the helix to be $\lambda = 2\pi a/\beta$.
The exchange field we use throughout is consistent with the parameters
discussed by Chiodi et al, 
\cite{chiodi}, as is the period of the spiral order along the c-axis.
While one could consider Tb or Dy~\cite{tbdy} or even MnSi \cite{mnsi}
which all exhibit
spiral magnetism, here we focus on Ho in which the proximity effect
has been more systematically established.

For notational convenience, Eq.~(\ref{exchange}) assumes a helix axis oriented along $\hat{z}$, i.e., parallel to the F/S interface as in Fig.~\ref{fig:1}; we also consider 
situations when this rotated by an angle $\theta\in[0,\pi]$ with respect to the $\hat{x}$-axis.   While in
an infinite system this axis direction is irrelevant,
when finite size effects are introduced it
is important as discussed below.
The second line in Eq.~(\ref{eq:full}) describes the pairing field 
$\Delta(\mathbf{r})=g(\mathbf{r})\langle\psi_{\uparrow}(\mathbf{r})\psi_{\downarrow}(\mathbf{r})\rangle$ 
of two fermions. 
This depends on the singlet pairing interaction $g(\mathbf{r})$, which is assumed to 
vanish in the ferromagnet and to be constant in the superconductor.
Although there is no intrinsic pairing in F, a singlet pairing correlation,
$\langle\psi_{\uparrow}(\mathbf{r})\psi_{\downarrow}(\mathbf{r})\rangle$
can be induced via proximity effects.

\section{Topological Features of Related Analytical Model}
To understand possible topological phases, 
it is useful to first neglect the position dependence in 
$\Delta(\mathbf{r})$ and focus on an infinite
superconductor with a uniform (non proximity induced) gap parameter $\Delta$; the magnitude of the
helical ferromagnet field strength is
assumed constant throughout. 
Applying a gauge transformation
$\psi_{\uparrow}\rightarrow e^{-i\beta z/2a}\psi_{\uparrow}$ and 
$\psi_{\downarrow}\rightarrow e^{i\beta z/2a}\psi_{\downarrow}$,
the single particle part of Eq.~(\ref{eq:full}) is
\begin{eqnarray}
H_{\rm sp}^\prime + H_{\rm FM}^\prime = H_{\rm sp} - v_{\rm so} \sigma^z i \partial_z + m v_{\rm so}^2/2+{\bf h}^\prime \cdot \boldsymbol{\sigma}  
\end{eqnarray}
where $\mathbf{h'}=h_0\left(\sin\alpha\mathbf{\hat x}+\cos\alpha\mathbf{\hat z}\right)$ and $v_{\rm so}=\beta/2 m a$.
In this way, the presence of helical magnetism
can be viewed as imposing the important combination of a constant Zeeman field
and one-dimensional spin-orbit coupling~\cite{spielmansoc,martin,Yaz3,JelenaK}.

Because the gauge transformed Hamiltonian is translationally invariant, we consider the Nambu spinor $\Psi_{\mathbf{k}} = \left( \psi_{\uparrow \mathbf{k}}, \psi_{\downarrow \mathbf{k}}, \psi^\dagger_{\uparrow \mathbf{k}}, \psi^\dagger_{\downarrow \mathbf{k}} \right)^T,$ allowing
 Eq.~(\ref{eq:full})
to be expressed in BdG form as
$\mathcal{H} = \frac{1}{2}\sum_{\mathbf k}\Psi_{\mathbf k}^{\dag}H_{\mathrm{eff}}\Psi_{\mathbf k}$, 
where
\begin{eqnarray}
\label{toy}
H_{\mathrm{eff}}&=&\epsilon_{\mathbf{k}}\tau^z - v_{\rm so} k_z\sigma^z+
h_0\left(\cos\alpha\sigma^z+\sin\alpha\sigma^x \right)\tau^z  \nonumber \\
& + & \Delta (i\sigma^y) \tau^{+} + \Delta^* (-i\sigma^y) \tau^{-} .
\end{eqnarray}
Here ${\bf k} = (k_x, k_y, k_x)$ is a 3D momentum for a dispersion $\epsilon(\mathbf k)=\mathbf k^2/2m-\mu$, where $E_F \to \mu = E_F - m v_{\rm so}^2 / 2$ is a renormalized chemical potential; in this work, we presume $\mu>0$ as is appropriate to F/S heterojunctions.
We also define the Pauli matrices $\tau^i$ to operate in particle-hole space with $\tau^\pm =\left(\tau^x+i\tau^y\right)/2$.

The topological phase diagram is analytically tractable for the conical opening angle $\alpha = \pi/2$. 
In a different context, this Hamiltonian has been explored elsewhere at other opening angles~\cite{1674-1056-24-5-050502,PhysRevA.93.063606,PhysRevLett.109.105303}.
The four bands of the quasi-particle energy spectrum satisfy
\begin{align}
E_{\bf k}^2 = \epsilon_{\bf k}^2+v_{\rm so}^2 k_z^2 + h_0^2+|\Delta|^2\pm2\sqrt{\epsilon_{\bf k}^2 \left(v_{\rm so}^2 k_z^2 + h_0^2 \right) + h_0^2|\Delta|^2}.
\end{align}
The one-dimensionality (or ``equal Rashba-Dresselhaus'' limit) of the spin-orbit coupling results in 
a topological phase structure that is qualitatively distinct from that frequently studied in proximity systems.
Rather than gapped topological phases, the above dispersion relation is associated with bulk nodal topological phases, which are
characterized~\cite{PhysRevLett.109.105303,ourTop}, 
by analyzing the spectrum as a function of $k_{\perp}\equiv\sqrt{k_x^2+k_y^2}$. The physics of the nodal points depends crucially 
on the dimension of the system, with a three-dimensional system having zero, one, or two nodal lines, 
while in two dimensions there are zero, two, or four Dirac points~\cite{PhysRevLett.109.105303}.

It is important to establish that these topological nodal features are robust.
We consider a perturbation of the form $m\tau^i\sigma^j$.
As long as particle-hole symmetry $\Xi =*\tau^x$ (where $*$ is anti-unitary complex conjugation)
and chiral symmetry $\mathcal C = -\tau^y\sigma^x$ are preserved, the only effect of $m$ 
is to renormalize $\mu, \Delta$ or $h_0$. Notably
perturbations of this form will not introduce a gap in the system.
This chiral symmetry is exact for $\alpha=\pi/2$, and therefore,
in this case, the nodal structure is
topologically protected. 

In particular we consider the 2D  limit by setting $k_y=0$; when $h_0<|\Delta|$, the system is gapped and in the trivial phase. 
When $|\Delta|< h_0<\sqrt{\mu^2+|\Delta|^2},$ 
the gap closes resulting in a topological phase with 
four Dirac points at $k_x^2 = \left( k_x^\pm \right)^2 \equiv 2m\left(\mu \pm \sqrt{h_0^2 - |\Delta|^2} \right)$. 
In the strong magnetic field limit, $h_0>\sqrt{\mu^2+|\Delta|^2}$,
the two Dirac points at $\pm k_x^{-}$ annihilate, resulting in two total Dirac points at $\pm k_x^{+}$.
The boundaries of these inequalities correspond to topological phase transitions.  

In a general topologically-non-trivial phase, one finds
gapless surface modes~\cite{RevModPhys.82.3045, RevModPhys.83.1057}; specifically, for topological superconductors
these are Majorana modes~\cite{Majorana, 2011arXiv1112.1950B, 0034-4885-75-7-076501}.
To analytically establish these Majorana surface modes we demonstrate a correspondence with
the well known Su-Schrieffer-Heeger
(SSH) \cite{PhysRevD.13.3398, PhysRevLett.42.1698} model.
Without loss of generality, we take $k_y=0$ and look at low energy excitations around the two nodal points
$\mathbf{k}^{\pm}_c=(\pm k_x^{+},0,0)$; a similar argument follows expanding around the points $\pm k_x^{-}$.
When the system has a finite extent, these will turn out to be connected by 
a flat band, as in the left panel of Fig.~\ref{fig:2}. 
Let us expand around the nodal points $\mathbf k=\mathbf{k}^{\pm}_c+\mathbf q$: to first order in $\mathbf q$, 
$H_{0}(\mathbf k) = H_{0}(\mathbf k_c)+\nabla_{\mathbf k}H_{0}(\mathbf k_c)\cdot\mathbf q$. The matrix $H_{0}(\mathbf k_c)$ has two non-zero eigenvalues along with two vanishing eigenvalues. Projecting into the degenerate subspace of these latter two eigenvalues 
yields 
\begin{equation}
\nabla_{\mathbf k} \bar{H}_{0}(\mathbf k_c)\cdot\mathbf{q} = \pm v_1 q_x \bar{\sigma}^1 + v_{\rm so} q_z  \bar{\sigma}^2
\label{eq:ssh}\end{equation}
where $v_1 = {\Omega k_x^{+}}/{mh_0}$.
Here $\Omega=\sqrt{h_0^2-\Delta^2}$ and the rotated Pauli matrices are given by 
$\bar{\sigma}^1 = \sigma^x$, $\bar{\sigma}^2 = (h_0 \sigma^{z} + i \Omega \sigma^{y}) / h_0$, 
and the two distinct signs $\pm$, reflect distinct fermion helicities.
Equation~(\ref{eq:ssh}) is evidently of the form of an effective
Su-Schrieffer-Heeger \cite{PhysRevD.13.3398, PhysRevLett.42.1698} Hamiltonian.
In particular, here we contemplate an interface at $z=0$ separating
different phases $q_x>0$ and $q_x<0$ so that  
this mapping establishes the existence of surface states at the $z=0$ interface. 
Importantly, these correspond to localized zero energy Majorana modes.
When we consider an extended system along $\hat{x}$, a value of $k_x^2 > 0$ will drive
a topological phase transition in the dimensionally reduced Hamiltonian. This results in a flat
band of Majorana edge states connecting the Dirac points.

\begin{figure}
\includegraphics[width=2.8in,clip]{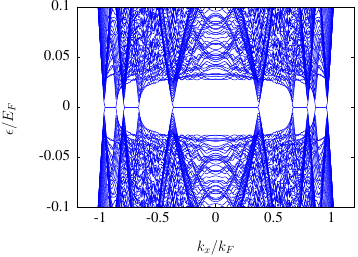}
\caption{The dispersion relation for a thin (homogenous) topological superconductor ($k_Fd_y=10$) where the exchange field strength and the pair potential are forced to be uniform throughtout the bulk. The three distinct zero-energy flat bands, and additional Dirac points, can be understood as coming from a finite-size quantization of modes along the $\hat{y}$-direction. The location of these flat bands is consistent with an analytic analysis of the BdG dispersion.
 \label{fig:ThinBands}}
\end{figure}

\subsection{Triplet Pairing Correlations} 
Motivated by interest from the superconducting
spintronics community, we address the triplet,
time dependent correlation functions
$f_{0}\left(\mathbf{r},t\right)= \left( f_{\uparrow\downarrow}\left(\mathbf{r}
,t\right)+f_{\uparrow\downarrow}\left(\mathbf{r},t\right) \right)/2$
and
$f_{1}\left(\mathbf{r},t\right)=\left( f_{\uparrow\uparrow}\left(\mathbf
{r},t\right)-f_{\downarrow\downarrow}\left(\mathbf{r},t\right) \right)/2$,
where $f_{\sigma\sigma^{\prime}}\left(\mathbf{r},t\right)=\left\langle
\psi_{\sigma}(\mathbf{r},t)\psi_{\sigma^{\prime}}(\mathbf{r},0)\right
\rangle
$.
Both components can be seen to vanish at $t=0$. Using Eq.~(\ref{toy})
one can calculate the anomalous Green's function $F_{\sigma\sigma'}(\omega_{n})$
as the Fourier transform of $f_{\sigma\sigma^\prime}(t)$, and 
it follows that the corresponding odd frequency pair amplitude defined through
$f_{\sigma\sigma'}^{-}(\omega_{n}) \equiv\frac{1}{2}[F_{\sigma\sigma'}(\omega_{n})-F_
{\sigma\sigma'}(-\omega_{n})]$
~\cite{PhysRevB.91.054518} satisfies
\begin{align}
f_{\uparrow\uparrow}^-=-f_{\downarrow\downarrow}^-=\frac{-2i\omega_nh_0\Delta}{\alpha(\omega_n,\mathbf k)},\ f_{\uparrow\downarrow}^-=f^-_{\downarrow\uparrow} = 0
\label{eq:8}
\end{align}
with $\alpha\left(\omega_{n},\mathbf{k}\right)$ even in $\mathbf{k}$
and Matsubara frequency $\omega_{n}$. In this way both the $m_{s}=1$ and $m_{s}=-1$
($f^-_{\uparrow\uparrow}$ and $f^-_{\downarrow\downarrow}$) triplet
correlations can be present while $m_{s}=0$
($f^-_{\uparrow\downarrow}+f^-_{\downarrow\uparrow}$)
is entirely absent. As a consequence, the $m_{s}=0$ component, if it
is present, can only be induced near the edge, where Majorana
modes appear.

\begin{figure}
\includegraphics[width=2.8in,clip] {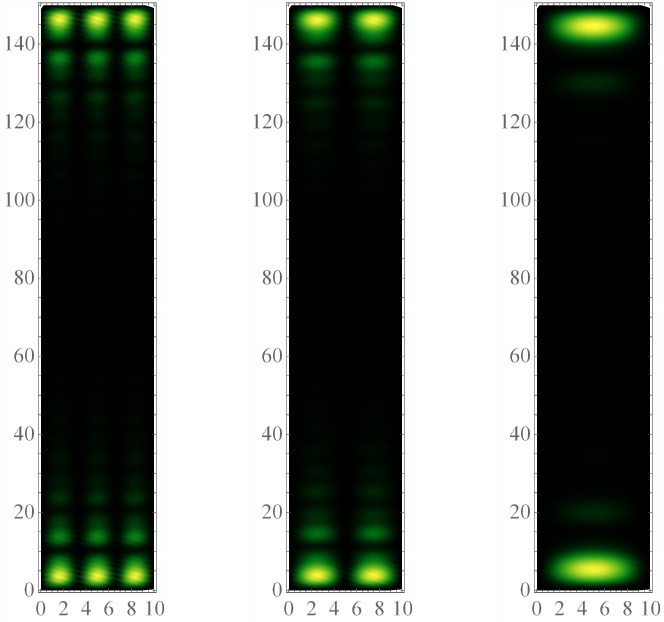}
\caption{BdG wavefunctions of a zero energy mode for the three flat bands observed in Fig.~\ref{fig:ThinBands}. From the left to the right panel, wavefunctions belong to the zero energy band of the range $0<k_x<0.37$, $0.66<k_x<0.79$, and $0.86<k_x<0.96$ are plotted respectively. The number of maxima along the $\hat{y}$- directions decreases when $k_x$ increases, consistent with our analytic analysis.\label{fig:ThinWF}}
\end{figure}

\begin{figure}
\includegraphics[width=2.8in,clip]{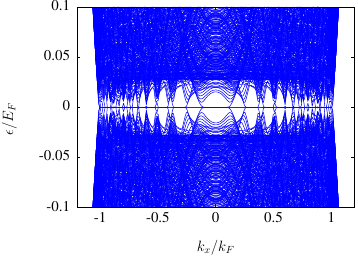}
\caption{The BdG dispersion relation for a wider junction, where $k_Fd_y=50$. The momentum modes $k_y=n\pi/k_Fd_y$, along $\hat{y}$, are still quantized, but with spacing that is five times denser than the thin junction discussed in Fig.~\ref{fig:ThinBands}. Therefore, there are many more flat bands correspond to each $k_y$ than the case shown in Fig.~\ref{fig:ThinBands}.
The numerous flat bands significantly overlap and approximately recover the two-dimensional flat band connecting Weyl rings for the case of a fully three-dimensional (nodal) topological superconductor. This is as would be expected from Eq.~(5).\label{fig:ThickBands}}
\end{figure}

\section{Numerical Study of Proximity Systems}
The above mapping onto the SSH model suggests that the proximity region of 
a F/S heterostructure can host interesting
topological phases with Majorana flat band edge states.
We next confirm this by numerically implementing the counterpart proximity junction
calculations using a fully
self-consistent scheme which allows for a spatially inhomogeneous
gap $\Delta({\bf r})$.
Specifically, we consider the geometry in Fig.~\ref{fig:1},
where, importantly, the system has finite extent along the junction direction $\hat{y}$,
as well as along the helical axis $\hat{z}$. We
consider the $x$ direction as infinite with a momentum labeled
by $k_x$.

The pair potential and exchange fields are functions of $y$ and $z$,
and
because the Hamiltonian $\mathcal{H}$ is  translationally invariant along $\hat{x}$,
this property allows us to write
\begin{align}
u_{n\sigma}({\bf r})&=\tilde{u}_{n\sigma}({\bf r}_\bot)e^{ik_xx},\\
v_{n\sigma}({\bf r})&=\tilde{v}_{n\sigma}({\bf r}_\bot)e^{ik_xx},
\end{align}
where ${\bf r}_\bot=(y,z)$ and $k_x$ is the momentum along $\hat{x}$ direction.
Then
for each value of $k_x$, 
we then solve for
the BdG eigenvalues and eigenfunctions

\begin{align}
\begin{pmatrix}
H_{\rm sp}(k_x,{\bf r}_\perp) + H_{\rm FM}({\bf r}_\perp) & \Delta({\bf r}_\perp) \left( i\sigma^y \right) \\
 - \Delta^*({\bf r}_\perp) \left( i\sigma^y \right) & - H_{\rm sp}^*(k_x,{\bf r}_\perp) - H_{\rm FM}^*({\bf r}_\perp)
\end{pmatrix}&\nonumber\\
\times
\begin{pmatrix}
\tilde{u}_{n\sigma}({\bf r}_\bot)\\
\tilde{v}_{n\sigma}({\bf r}_\bot)
\end{pmatrix}
=\epsilon_n
\begin{pmatrix}
\tilde{u}_{n\sigma}({\bf r}_\bot)\\
\tilde{v}_{n\sigma}({\bf r}_\bot)
\end{pmatrix}.&
\label{eq:11}
\end{align}

Here, the momentum label $k_x$ only enters in $H_{\rm sp}(k_x,{\bf r}_\perp) = k_x^2/2m - \nabla^2_\perp/2m - \mu$. This acts to shift the chemical potential $\mu \rightarrow \mu - k_x^2/2m$, and in this way, the system is ``dimensionally reduced'' with respect to the topological properties.

The self-consistent order parameter (which depends on the
interaction strength 
$g(\mathbf{r})$) 
is to be
distinguished from the anomalous pairing amplitudes. The former is
zero in the ferromagnet, while the latter is not. We have
\begin{equation}
\Delta(\mathbf {r})=g(\mathbf{r})\langle\psi_{\uparrow}(\mathbf{r})\psi_{\downarrow}(\mathbf{r})\rangle \equiv g(\mathbf{r})F(\mathbf{r})=g({\bf r}_\bot)F({\bf r}_\bot).
\end{equation}
We similarly define the pair amplitudes $F(\mathbf{r})\equiv\langle\psi_{\uparrow}(\mathbf{r})\psi_{\downarrow}(\mathbf{r})\rangle$.

\subsection{Introducing the Helical magnet}

The exchange field $\mathbf{h}({\bf r}_\bot)$ of the ferromagnet is given 
in Eq.~(\ref{exchange}), 
and is taken to vanish in the superconductor and to be present in the ferromagnet, so that $h_0 ({\bf r}_\bot) = h_0 \Theta(d_F - y)$.
For most of this paper, the helical axis of the exchange field $\mathbf{h}$ is
along the $\hat{z}$ direction. However, in experimental junctions
one can contemplate a more general expression for $\mathbf{h}$ with the helical axis
lying in the $y-z$ using the rotation matrix,
\begin{equation}
R_x(\theta)=
\begin{pmatrix}
 1&0          &0         \\
 0&\cos\theta &\sin\theta\\
 0&-\sin\theta&\cos\theta
\end{pmatrix}.
\end{equation}
As a consequence the rotated exchange field $\mathbf{h}(\mathbf{r})$ is 
\begin{equation}
\mathbf{h}(\mathbf{r})\rightarrow \mathbf{\bar{h}}(\mathbf{r}) = R_x(\theta)\mathbf{h}\left(R_x^{-1}(\theta)\mathbf{r}\right).
\end{equation}
The helical axis of $\mathbf{\bar{h}}(\mathbf{r})$ now makes an angle $\theta$ with respect to the $\hat{z}$-axis.
This leads to 
\begin{widetext}
\begin{eqnarray}
\bar{h}_{x} & = & h_{0}\sin\alpha\left[\cos\left(\frac{\beta}{a}\sin\theta y\right)\cos\left(\frac{\beta}{a}\cos\theta z\right)-\sin\left(\frac{\beta}{a}\sin\theta y\right)\sin\left(\frac{\beta}{a}\cos\theta z\right)\right],\\
\bar{h}_{y} & = & h_{0}\cos\theta\sin\alpha\left[\sin\left(\frac{\beta}{a}\sin\theta y\right)\cos\left(\frac{\beta}{a}\cos\theta z\right)+\cos\left(\frac{\beta}{a}\sin\theta y\right)\sin\left(\frac{\beta}{a}\cos\theta z\right)\right]+h_{0}\sin\theta\cos\alpha,\\
\bar{h}_{z} & = & -h_{0}\sin\theta\sin\alpha\left[\sin\left(\frac{\beta}{a}\sin\theta y\right)\cos\left(\frac{\beta}{a}\cos\theta z\right)+\cos\left(\frac{\beta}{a}\sin\theta y\right)\sin\left(\frac{\beta}{a}\cos\theta z\right)\right]+h_{0}\cos\theta\cos\alpha.
\end{eqnarray}
\end{widetext}

\begin{figure}[h]
\includegraphics[width=2.8in,clip]{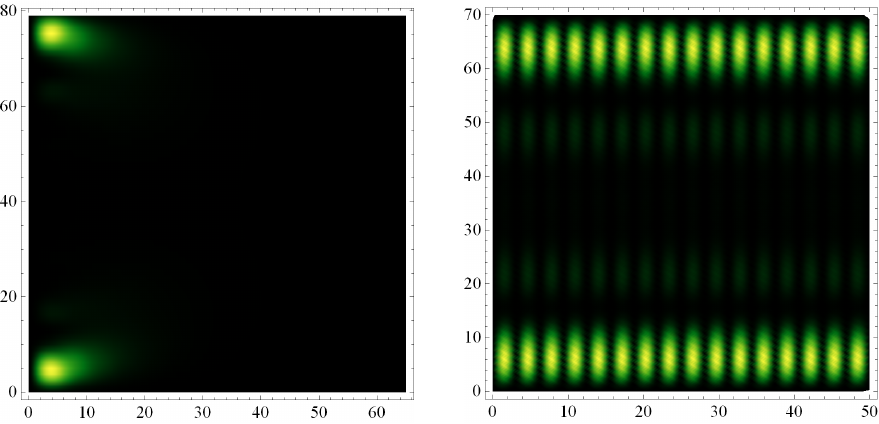}
\caption{
The BdG wavefunction of a Majorana zero mode for the (a) proximity-coupled junction
and (b) a homogeneous system. (a) A calculation using the self-consistent gap
in the geometry and parameters of Fig.~2(a) of the main text, with $k_x$ lying in the flat band.
(b) The zero mode wavefunction with the gap and exchange field are
forced to be homogeneous throughout the sample. This calculation helps disentangle the role of
finite size effects from the inhomogeneous gap in the self-consistent calculation.
Here, we take $h_0=0.1E_F$, and $k_x=0$ to lie in the central flat band (see also Fig.~\ref{fig:ThinBands});
all other parameters are the same as the proximity-coupled junctions.
The nodal structure in the BdG wavefunction demonstrates a 3D like nature,
in contrast to the quasi-2D like nature on the left panel.
We find the wavenumber is approximately $k_F$, as is expected from the discussion of the analytic model.}
\label{fig:1s}
\end{figure}

\begin{figure*}
\includegraphics
[width=5.3in]
{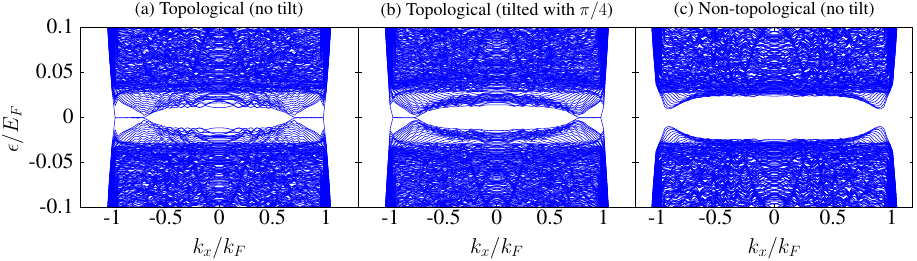}
\caption{
The dispersion of the BdG Hamiltonian (see main text) as a function of $k_x$ for three different helicial ferromagnets.
(a) The helical axis is oriented along $\hat{z}$: zero-energy flat bands connect two pairs of topologically protected Dirac points, as predicted from the analytic model.
(b) Rotating the helical axes by $\pi/4$ about the $\hat{x}$-axis, we find the Dirac cones and edge states remain.
(c) With decreased ferromagnetic field strength
and without rotating the axis, the system enters a trival phase.
}
\label{fig:2}
\end{figure*}

\subsection{Numerical algorithm}

We numerically solve the BdG eigenvalue problem of Eq.~({\ref{eq:11})
following the scheme developed in Ref.~\cite{halterman2}.
For definiteness,  we set the lattice constant to be $a=k_F^{-1}$.
We then expand both the matrix elements of Eq.~(\ref{eq:11}) and
the eigenfunctions in terms of a Fourier basis.
For the quasi-particle and quasi-hole wavefunctions, we have
\begin{eqnarray}
\tilde{u}_{n\sigma}({\bf r}_\bot)&=&\frac{2}{\sqrt{d_yd_z}}\sum_{p,q}u_{n\sigma}^{pq}\sin\left(\frac{p\pi y}{d_y}\right)\sin\left(\frac{q\pi z}{d_z}\right),\\
\tilde{v}_{n\sigma}({\bf r}_\bot)&=&\frac{2}{\sqrt{d_yd_z}}\sum_{p,q}v_{n\sigma}^{pq}\sin\left(\frac{p\pi y}{d_y}\right)\sin\left(\frac{q\pi z}{d_z}\right).
\end{eqnarray}
We generically define the matrix elements of an operator $M$ to be
\begin{eqnarray}
M^{pqp'q'} & \equiv & \langle pq|M|p'q'\rangle \\
& =&
\frac{4}{d_{y}d_{z}}\int_{0}^{d_{y}}\int_{0}^{d_{z}}dydz \sin\left(\frac{p\pi y}{d_{y}}\right)\sin\left(\frac{q\pi z}{d_{z}}\right)\nonumber  \times \\
& &
 \qquad M \sin\left(\frac{p'\pi y}{d_{y}}\right)\sin\left(\frac{q'\pi z}{d_{z}}\right).
\end{eqnarray}

Our BdG eigenvectors are then used to construct a self-consistent gap profile
\begin{equation}
\Delta(\mathbf{r}_\perp) = g(\mathbf{r}_\perp) \sum_{\epsilon_n<\omega_D} \left(u_{n\uparrow}v_{n\downarrow}^*-u_{n\downarrow}v_{n\uparrow}^*\right) \tanh(\frac{\epsilon_n}{2T}),
\end{equation}
where the Debye frequency $\omega_D$ is the energy cutoff and $T$ is the temperature (we set $T=0$ in this paper).
Similarly
the pairing amplitudes
are found to be
\begin{equation}
F(\mathbf{r}_\perp) = \sum_{\epsilon_n<\omega_D} \left(u_{n\uparrow}v_{n\downarrow}^*-u_{n\downarrow}v_{n\uparrow}^*\right) \tanh(\frac{\epsilon_n}{2T}),
\end{equation}
where the sum over the energy index $n$ also includes an integral over $k_z$ states.
The coupling function $g({\bf r}_\bot)=g \Theta(y-d_F)$, where $\Theta(y)$ is the unit step function, is taken to be a constant
inside the superconducting region while vanishing in the ferromagnet.

Where topological phases enter is governed by the details of the
resulting energy dispersion in the Ho subsystem. These, in turn,
depend on the pairing
correlations.
These correlations are associated with
real space pairing oscillations (deriving from LOFF-like physics) 
and depend rather strongly on $h_0$. In
this context, and because of the inequalities associated
with topological order, the value presumed for $h_0$ is important,
and is here taken to agree with experiment \cite{chiodi}. 
Another parameter which could be of
concern is the energy cutoff $\omega_D$.
This
sets the overall superconducting transition temperatures
of the pure bulk superconductor~\cite{tinkham}, but is otherwise irrelevant
when discussing topological inequalities. For definiteness,
we take $\omega_D=0.1E_F$
in all of our numerical calculations.

With this analysis, we are able to transform the inhomogenous BdG differential equation into an algebraic matrix that can be numerically diagonalized~\cite{halterman2} to produce eigenvalues 
($\epsilon_n$)
and eigenfunctions ($u_n$ and $v_n$) thereby obtaining 
essentially all important quantities~\cite{halterman2,supp}.

\section{Finite Size Effects in Homogeneous (Non-Proximity) Superconductors}
In order to calibrate the results obtained in a proximity
junction, it is useful first to identify signatures of Majorana phases in
a simpler situation in which a true order parameter is presumed as in
the analytical model, but in a system with
finite sample dimensions. This situation is
more physical than in the analytically tractable model
since these finite dimensions are inevitable
in a proximity situation. Here,  we perform a series
of numerical calculations based on the assumption of a homogeneous
gap taking the exchange field and
the pair potential as constant. The analysis here builds on
the numerical algorithm developed in the previous section,
without the complexity of establishing a self-consistent
spatially dependent gap.
By varying the sample thickness systematically, in effect, we study a crossover of the edge mode structure from quasi-2D to 3D.
Several different thicknesses $k_F d_z$ were considered, ranging from $k_F d_z=70$ to $k_F d_z=150$,
and our results were found to be independent of this parameter for all the cases considered.
In all calculations we assumed that the sample was infinite in the $\hat{x}$-direction.

We can more carefully map out how the system evolves 
by analyzing the behavior of the
energy dispersion.
The positive-energy branches of the continuum BdG dispersion are
\begin{equation}
E_{\bf k}=\sqrt{\epsilon_{\bf k}^2+v_{\rm so}^2 k_z^2 + h_0^2+|\Delta|^2\pm2\sqrt{\epsilon_{\bf k}^2 \left(v_{\rm so}^2 k_z^2 + h_0^2 \right) + h_0^2|\Delta|^2}}.
\end{equation}
For a 3D homogeneous and infinite sample, nodal points can only occur when $k_{z}=0$, which gives
\begin{equation}
E(k_{z}=0)=\sqrt{\left(k_{x}^{2}+k_{y}^{2}-1+mv_{\rm so}^2/2 E_F\right)^{2}+\Delta^{2}}-h_{0},
\end{equation}
where we have chosen the negative sign. For simplicity, here and in the remainder of this section,
we normalize energies to $E_F$ and momentum to the Fermi wavevector, $k_F$. 
The correction to the chemical potential by the helical wavevector $\beta=\frac{\pi}{6a}$ is $\mu - E_F = mv_{\rm so}^2/2 \approx 0.07E_F$  (see main text).

We consider a finite sample of width $d_{y}$ in the $\hat{y}$ direction, with
$k_{y}^{\min}=\frac{\pi}{k_{F}d_{y}}$ and $k_{y}^{\max}\approx1$.
When $d_{y}$ goes to infinity, $k_{y}^{\min}$
vanishes. However, for a sufficiently small sample, the discretized nature will result in quantized momentum modes $k_{y}=\frac{n\pi}{k_{F}d_{y}}$ for integer $n$. The resulting excitation spectrum will be approximated by ``cuts'' through the full 3D spectrum, where $k_y$ is fixed.
To see this, consider the spectrum when $k_y$ has been replaced with a quantized mode:
\begin{equation}
E(k_{z}=0)=\sqrt{\left(k_{x}^{2}+\left(\frac{n\pi}{k_{F}d_{y}}\right)^2-1+mv_{\rm so}^2/2 E_F\right)^{2}+\Delta^{2}}-h_{0}.
\end{equation}
This quantity vanishes when
\begin{equation}
k_{x}^{2}=1-mv_{\rm so}^2/2 E_F-\left(\frac{n\pi}{k_{F}d_{y}}\right)^{2}\pm\sqrt{h_{0}^{2}-\Delta^{2}}.
\end{equation}
is satisfied. As a result, the finite size along $d_y$ effectively renormalizes the chemical potential and shifts the topological phase boundary.

For the example we address below, we take $h_{0}=0.1$, $\Delta=0.032$,
$k_{F}d_{y}=10$ and $k_{F}d_{z}=150.$ Thus the two nodal points occur at $k_{x}^{-}=0.86$ and $k_{x}^{+}=0.96$. The zero energy flat band in a finite-size system then lies in the range $0.86<k_{x}<0.96$.
When $n=2,$ we have $0.66<k_{x}<0.79$. For $n=3,$ we have $0<k_{x}<0.37$.
This analysis shows that there will be three zero energy flat bands
for $n=1,2,3$. 
Furthermore, for each band indexed by
$n$, the corresponding wavefunction amplitudes show $n$ maxima along
the $\hat{y}$-direction at two edges in the $\hat{z}$-direction. This is
illustrated as Fig.~\ref{fig:ThinWF}.

To understand the limit as the width $d_{y}$ tends to infinity, we take the case of
a large but finite  $y$-thickness ($k_{F}d_{y}=50$), as illustrated in the
electronic structure in Fig.~\ref{fig:ThickBands}. One can understand the limiting case by first
visualizing a Weyl annulus lying on $k_{x}-k_{y}$ plane. If one looks
at the annulus along the $k_{y}-$axis, the Weyl annulus becomes a complete
line. The plot in 
Fig.~\ref{fig:ThickBands} should be contrasted with that in
Fig.~\ref{fig:ThinBands}.

\begin{figure*}
\includegraphics[width=4.9in,clip]{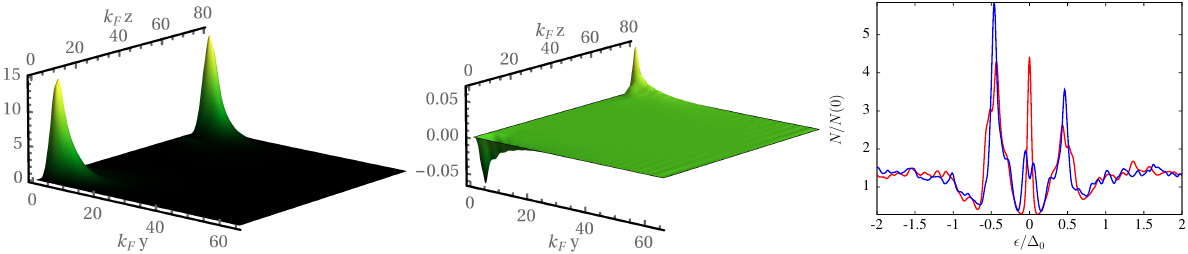}
\caption{
Plots of the (a) BdG wavefunction of the zero energy mode at $k_x=0.95 k_F$ and (b) $m_s=0$ triplet correlations
as measured by the real part of $f_0(t=4/\omega_D)$ (the magnitude of the imaginary part is much smaller.)
These triplet correlations cannot appear in the bulk, and only
exist near the edge for both topological and trivial phases.
(c) The local density of states (DOS), normalized to the DOS of the normal metal at the Fermi-level.
For a thicker sample size, ($d_z=80/k_F$, red) a single zero-energy peak is observed, corresponding to localization of Majorana edge states.
As the thickness is decreased ($d_z=30/k_F$, blue), the Majorana states can overlap, resuling in a splitting of the zero-energy state peak.
}
\label{fig:3}
\end{figure*}

\section{Numerical evidence for Majorana flat bands in proximity junctions}
With this framework we now present numerical solutions to the BdG equations 
in a proximity configuration, as given in Eq.~(\ref{eq:11}); this shows how
Majorana modes discussed above will appear in a conical ferromagnet-
superconductor junction.
Unless otherwise specified, our geometry is based upon a superconducting (S) region
with a coherence length $\xi = 20/k_F$.
This S region is large, with thickness (in the $\hat{y}$ direction) $60/k_F$, 
whereas we consider widths of the ferromagnet varying from $5/k_F$ to $20/k_F$.
The height of the F/S junction in the $\hat{z}$-direction is $80/k_F$.
The exchange field in Figs.~\ref{fig:2}(a) and~\ref{fig:2}(b) is $h_0 = 0.2 E_F$.

We first observe from Figs.~\ref{fig:1}(b)
and \ref{fig:1}(c)
that the pairing amplitude
penetrates significantly into the F region which is to the 
left of the vertical dashed lines.
Figure
\ref{fig:1}(b)
corresponds to the case when
the thickness of F is comparable to the coherence length. Here
one sees that the
singlet pairing amplitude exhibits oscillations of a LOFF-like form.
As shown in this figure, when the oscillation length (which depends on the exchange field)
is shorter than the width of the ferromagnetic region,
the singlet pairing amplitude can reach zero sufficiently deep into the conical magnet. Consequently,
the bulk energy spectrum is no longer fully gapped and this will destroy
a topological phase..
We therefore chose the width of the F region to be large compared to the Fermi length,
but still sufficiently small such that the singlet component does not
assume a value of zero in the bulk.
This is is the situation shown in
Fig.~\ref{fig:1}(c), which is the basis for our subsequent analysis.

As a summary figure, it is useful to
first to compare results for
a true proximity system with one in which there is
a homogeneous gap, as in the figures of the previous section.
Figure~\ref{fig:1s} compares an (a) edge mode found in the proximity coupled system to (b) that of a system with
a homogeneous gap throughout. In the proximity calculation, the width $k_Fd_y=6$ of the ferromagnet is small
compared to the correlation length, and the edge mode is tightly localized in the proximity region. In contrast, the system with a homogeneous gap has an edge mode localized through the entire region of length $k_Fd_y=50$. This edge-state wavefunction has a nodal character signifying the wavefunction is a standing mode with wavevector close to $k_F$.

We turn to
Fig.~(\ref{fig:2}) which illustrates the
topological structure, and emergence of Majorana flat bands in
this proximity system.
Just as in the analytic model of Eq.~(\ref{toy})
there exists a critical value of $h_0$ that defines
a transition between a fully gapped trivial phase
and a topological phase with two or four Dirac points.
We find that the numerically
determined topological phase boundaries
do not match precisely with those of the analytical
model. This is presumably because there is no well defined
value to assign to the parameter, $\Delta$, since
the coupling $g(\mathbf{r}_\perp)$ is zero in the ferromagnet; while
there is a pairing gap which is proximity induced, it
assumes a range of position dependent values.

In Fig.~\ref{fig:2}
we provide examples of the calculated energy dispersion for different orientations of the helical axis and in topological and trivial phases.
As above, we fix the opening angle $\alpha=\pi/2$.
The helical axes in Figs.~\ref{fig:2}(a) and \ref{fig:2}(c) are along
${\mathbf{\hat z}}$, while it is along
${\mathbf{\hat y}}+{\mathbf{\hat z}}$,
in Fig.~\ref{fig:2}(b).
We see from the central panel that,
as expected, robust Majorana phases
survive up to some reasonably large rotation angle, illustrated here
with $\pi/4$.  For the case of a $\pi/2$ rotation
additional complications, relating to LOFF oscillations ensue.
By contrast, Fig.~\ref{fig:2}(c) presents
the case of a smaller field strength $h_0 = 0.08E_F$,
such that the system
has crossed from the topological phase shown in the
first panel, into the trivial phase.
This results
in a superconductor that is fully gapped without surface states.

To establish whether the zero energy flat bands of
Fig.~\ref{fig:2} are
related to surface Majorana effects,
in Fig.~\ref{fig:3}(a)
we study the localization of the flat-band eigenfunctions
in the $\hat{y}-\hat{z}$ plane.
We plot these wavefunctions taking
$k_x=0.95 k_F$ close to a bulk nodal point;
this wavefunction is localized near the $\hat{z}$-axis edge.
Notably it is
rather sharply peaked on the ferromagnetic side
away from the precise F/S interface.
This observation is what we would expect according to
the analytic discussion surrounding Eq.~(\ref{toy}).
This provides some support for the conclusion
that our numerical calculations have, indeed,
identified Majorana modes.

Before looking for additional support,
it is useful to see if these Majorana flat bands are related
to the
odd-frequency pairing amplitude, as has been suggested or 
investigated \cite{PhysRevB.83.224511, PhysRevB.86.144506, PhysRevB.87.104513, PhysRevB.91.054518,Galitskinew}.
In Fig.~\ref{fig:3}(b), we plot the
$m_s=0$ triplet correlation function $f_0$.
As shown in our analytical analysis of Eq.~(\ref{eq:8}) we should not
find a non-vanishing $m_s = 0$ triplet component  in the bulk.
This panel, indeed, shows that this particular correlation
is confined to the surface, much like the Majorana modes.
Notably, we find this to be the case even in the non-topological
phases, so that despite the fact that they appear rather similar
there is little direct correlation between the
$f_0$ and the Majorana modes.

To establish experimental signatures for the existence of Majorana modes,
we address the local density of states (LDOS) which can be probed using
scanning tunneling microscopy or photoemission.
Recent work~\cite{Marcus} exploring the nature of topological protection with end-mode separation,
suggests that we analyze how a finite  thickness in the $\hat{z}$ direction
affects the localized nature of our flat band, zero energy modes.
In the right panel of Fig.~\ref{fig:3} we plot
the LDOS for two different thicknesses.
The red curve corresponds to the thicker system with
$d_z=80/k_F$, where there is a single zero bias peak. For
the blue curve where the thickness is substantially
reduced, $d_z=30/k_F$, we find two peaks in the LDOS. This is
what would be expected if the surface states were Majorana modes
which had some overlap, due to finite size effects.

\section{Conclusion}

In this paper we have suggested a different heterostructure
for readily observing nodal topological superconductivity
and related Majorana surface flat bands. This is to be contrasted
with the widely studied
$p_x \pm i p_y$ proximity heterostructures~\cite{fu,sau} which yield a strictly
two or one dimensional (gapped) topological superconductor.
We consider proximity induced superconductivity in conical ferromagnets where
the necessary ingredients of effective (1D) spin-orbit coupling and Zeeman
fields are conveniently and simultaneously present. The feasibility of
making these superconducting heterostructures is
well established for Nb-Ho proximity junctions, where there is
clear evidence~\cite{wu1,robinson,linder}
for finite range penetration of superconducting correlations. Here, however,
we suggest that the helical axis of Ho be oriented parallel
to the junction plane.

We employ a numerical Bogoliubov-de Gennes scheme which can
accomodate finite length scales in two of the three dimensions.
While we encounter increased numerical complexity, in contrast to periodic boundary conditions, 
we then have
access to edges and can study the Majorana wavefunctions
associated with the $E=0$ flat bands.
This numerical scheme should also
be compared with Eilenberger-based approaches
which it was suggested \cite{Galitskinew} might be required in
order to establish triplet, odd frequency pairing.
We have demonstrated that such a pairing correlation is indeed found
in our BdG approach, and is associated with the edges, rather like
that of the $E=0$ Majorana states.
However, because these triplet effects also appear in a non-topological
phase, there is no simple correlation between the two.

We emphasize, throughout, that with our self-consistent proximity
calculation, there is a clear distinction between analytical models
for the equivalent topological phases (with a presumed homogeneous
order parameter) and the counterpart
proximity-induced
phase which contains no pairing interaction and thus no order
parameter in the magnetic subsystem. Nevertheless, 
a numerical study of a homogeneous system which includes
edges and finite size effects, provides a calibration showing
how Majorana flat band states will appear without proximity
effects. A comparison of the wavefunction plots we find in
our junctions (induced solely by proximity) provides strong
support for identifying these bound surface states
with 
Majorana flat bands.

Importantly, our
calculations show that localized Majorana
modes appear away from the junction interface.
This presents an experimental advantage 
as these Majorana states are much more accessible than in
a buried proximity interface.
In a rather complete review of intrinsic nodal topological superfluids
\cite{Schnyder}
(such as high $T_c$ cuprates, heavy fermions and the A
phase of helium-3), it was noted that 
the most reliable experimental signatures
of Majorana flat bands involve the tunneling conductance: in
particular a sharp zero bias peak. There are also weaker
indications in the electromagnetic response and the anomalous spin Hall
conductance and in quasi-particle interference using
scanning tunneling microscopy \cite{Schnyder}.
Additional interest has focused on
the anomalous Josephson effect \cite{AnomJosephson}.

In this paper we have singled out the zero bias tunneling feature and moreover
demonstrated how
it is modified as the height of the junctions is reduced. This
latter is suggestive of the
interaction between Majorana bound states~\cite{Marcus} and should serve
to more clearly identify these topological signatures.
We observe as well, that 
because the Majorana modes are at the
junction corner, they may be more amenable to photoemission probes.

We end by noting that we have not considered the effect of disorder
in our proximity calculations, although the $s$-wave
odd-frequency spin triplet state is not particularly
sensitive to impurity effects, as compared with the
$p$-wave even-frequency spin triplet~\cite{bergeret}. However, even for a pristine sample,
disorder is
inevitable at the surface. Our numerical scheme can and will be
extended to address these disorder effects in a future work.

\textit{Acknowledgements.--}
We thank Rufus Boyack for helpful conversations. This work was supported by NSF-DMR-MRSEC 1420709.

\appendix
\numberwithin{equation}{section}
\section{Additional Numerical Details on Proximity Calculations}

We present the expansion coefficients of the single particle part $H_{\mathrm{sp}}$, which can be found from explicit calculation to be
\begin{align}
H_{\mathrm{sp}}^{pqp'q'}=\left\{ \frac{1}{2m}\left[\left(\frac{p\pi}{d_{y}}\right)^{2}+\left(\frac{q\pi}{d_{z}}\right)^{2}+k_{x}^{2}\right]-E_{F}\right\} \delta_{pp'}\delta_{qq'}.
\end{align}
Similarly, the pair potential can also be expanded as
$\Delta^{pqp'q'} \equiv \langle pq|\Delta ({\bf r}_\bot) |p'q' \rangle $.
We also calculate the expansion coefficients for each vector component of exchange fields $h_i^{pp'qq'}$ as $\langle pq | h_i | p'q' \rangle$ $(i=x,y,z)$.
for general $\theta$ defining the angle between the z direction and
helical axes, we have
\begin{align}
h_x^{pp'qq'} &= h_{0}\sin\alpha\left(K_{1pp'}R_{1qq'}-K_{2pp'}R_{2qq'}\right),\\
h_y^{pp'qq'} &= h_{0}\cos\theta\sin\alpha\left(K_{2pp'}R_{1qq'}+K_{1pp'}R_{2qq'}\right)\\\nonumber
&+h_{0}\sin\theta\cos\alpha D_{pp'qq'},\\
h_z^{pp'qq'} & = -h_{0}\sin\theta\sin\alpha\left(K_{2pp'}R_{1qq'}+K_{1pp'}R_{2qq'}\right)\\\nonumber
&+h_{0}\cos\theta\cos\alpha D_{pp'qq'},
\end{align}
where we have defined the following quantities

\begin{align*}
K_{1pp'}^{\pm\pm} = & \frac{\sin\left(d_{F}\left(\frac{\beta}{a}\sin\theta\pm\frac{p\pm p'}{d_{y}}\pi\right)\right)}{\frac{\beta}{a}\sin\theta d_{y}\pm(p\pm p')\pi},\\
K_{2pp'}^{\pm\pm} = & \frac{1-\cos\left(d_{F}\left(\frac{\beta}{a}\sin\theta\pm\frac{p\pm p'}{d_{y}}\pi\right)\right)}{\frac{\beta}{a}\sin\theta d_{y}\pm(p\pm p')\pi},\\
R_{1pp'}^{\pm\pm} = & \frac{\sin\left(d_{z}\frac{\beta}{a}\cos\theta\pm\left(q\pm q'\right)\pi\right)}{d_{z}\frac{\beta}{a}\cos\theta\pm\left(q\pm q'\right)\pi},\\
R_{2pp'}^{\pm\pm} = & \frac{1-\cos\left(d_{z}\frac{\beta}{a}\cos\theta\pm\left(q\pm q'\right)\pi\right)}{d_{z}\frac{\beta}{a}\cos\theta\pm\left(q\pm q'\right)\pi},\\
D_{pp'qq'} = & \left[\frac{\sin\left(\frac{\left(p-p'\right)\pi d_{F}}{d_{y}}\right)}{\left(p-p'\right)\pi}-\frac{\sin\left(\frac{\left(p+p'\right)\pi d_{F}}{d_{y}}\right)}{\left(p+p'\right)\pi}\right]\delta_{qq'},
\end{align*}
and
\begin{eqnarray*}
K_{1pp'} & = & \frac{1}{2}\left(K_{1pp'}^{+-}+K_{1pp'}^{--}-K_{1pp'}^{++}-K_{1pp'}^{-+}\right),\\
K_{2pp'} & = & \frac{1}{2}\left(K_{2pp'}^{+-}+K_{2pp'}^{--}-K_{2pp'}^{++}-K_{2pp'}^{-+}\right),\\
R_{1pp'} & = & \frac{1}{2}\left(R_{1pp'}^{+-}+R_{1pp'}^{--}-R_{1pp'}^{++}-R_{1pp'}^{-+}\right),\\
R_{2pp'} & = & \frac{1}{2}\left(R_{2pp'}^{+-}+R_{2pp'}^{--}-R_{2pp'}^{++}-R_{2pp'}^{-+}\right).
\end{eqnarray*}
The differential BdG eigenvalue equations are now converted
into algebraic eigenvalue problems. The number of terms in the Fourier
series
is determined through the following relations
\begin{equation}
\frac{1}{2m}\left(\frac{n_{y,z}^{\rm max}\pi}{d_{y,z}}\right)^2=E_F(1+\omega_D).
\end{equation}
For numerical purposes, one has to use a finite number of points for the continuous variable ``$k_x$''.
We choose $k_x$ to be evenly distributed in the range $-k_x^{\rm max}<k_x<k_x^{\rm max}$, where
$k_x^{\rm max}$ is given by
\begin{equation}
\frac{\left(k_x^{\rm max}\right)^2}{2m}=E_F(1+\omega_D).
\end{equation}
We choose the number of $k_x$ points to be 256.

In order to find
the correct energy minimum, the singlet pair amplitudes are determined
self-consistently. In other words, the eigenvalue problem is solved iteratively.
Once we have the self-consistent quasi-particle eigenfunctions, the triplet amplitudes can be obtained via the following equations
\begin{align}
 & f_{0}({\bf r}_\bot,t)\nonumber\\
 & = \sum_{n}\left[u_{n\uparrow}({\bf r}_\bot)v_{n\downarrow}^{*}({\bf r}_\bot)+u_{n\downarrow}({\bf r}_\bot)v_{n\uparrow}^{*}({\bf r}_\bot)\right]\zeta_{n}(t),\\
 & f_{1}({\bf r}_\bot,t)\nonumber\\
 & = \sum_{n}\left[u_{n\uparrow}({\bf r}_\bot)v_{n\uparrow}^{*}({\bf r}_\bot)-u_{n\downarrow}({\bf r}_\bot)v_{n\downarrow}^{*}({\bf r}_\bot)\right]\zeta_{n}(t),
\end{align}
where $\zeta_{n}(t)=\cos\left(\epsilon_{n}t\right)-i\sin(\epsilon_{n}t)\tanh\left(\epsilon_{n}/2T\right)$.
Another quantity discussed in the main text is the local
density of states (LDOS). It is defined as
\begin{equation}
N_{\sigma}({\bf r}_\bot,\epsilon)=\sum_{n}\left|u_{n\sigma}\left({\bf r}_\bot\right)\right|^{2}\delta(\epsilon-\epsilon_{n})+\left|v_{n\sigma}\left({\bf r}_\bot\right)\right|^{2}\delta(\epsilon+\epsilon_{n}),
\end{equation}
where $\sigma=\uparrow,\downarrow$.

\section{Correspondence with Su-Schrieffer-Heeger Model in Case of Homogeneous Gap}

Here we provide details showing the relation to the SSH model 
\cite{PhysRevD.13.3398, PhysRevLett.42.1698}.

We use an expansion around the appropriate Dirac points in the effective model to suggest the existence of protected Majorana bound states.
In a topologically non-trivial phase, bulk topological signatures are associated with gapless surface modes. Specifically, we are considering a topological superconductor, where we expect the edge modes to be zero-energy Majorana states.
When $h_0>\Delta$ and $\Omega^2 \equiv h_0^2-\mu^2 < \Delta^2$, phase transitions are signaled by a closing BdG spectral gap at $\mathbf k=(\pm k_{x}^{\pm}, 0, 0)$, where
\begin{equation}
\frac{(k_x^{\pm})^2}{2m}-\mu = \pm\sqrt{h_0^2-|\Delta|^2}=\pm\Omega.
\end{equation}
Given a momentum $k_x$ in a non-trivial region $k_x^-<|k_x|<k_x^+$, there exists a surface state associated with a flat band connecting
two nodal points. We may show this explicitly by looking at low energy excitations around $k_x^{\pm}$.

The structure of the transition can be understood from a small-momentum expansion $\mathbf k=\mathbf k_c+\mathbf q$ around the point $\mathbf k_c=(\pm k_{x}^{+},0,0)$. To first order in $\mathbf q$, $H(\mathbf k) = H(\mathbf k_c)+\nabla_{\mathbf k}H(\mathbf k_c)\cdot\mathbf q$ with
\begin{eqnarray}
H(\mathbf k_c) & = & \Omega\tau^z+h_0\tau^z\sigma^x+\tau^+(i\sigma^y)\Delta+\mathrm{h.c.},\\
\nabla_{\mathbf k}H(\mathbf k_c) & = & \pm\hat{\mathbf x}\frac{k_x^{+}}{m}\tau^z-\hat{\mathbf z}v_{\rm so}\sigma^z.,
\end{eqnarray}

Our BdG Hamiltonian has four distinct energy bands due to broken spin degeneracy, two are particle-like and two hole-like. The spectrum is gapless  at $\mathbf k_c$ in the sense that the lower particle band and the higher hole band vanish at this point. This can be understood explicitly by diagonalizing $H(\mathbf k_c)\to O^{-1}H(\mathbf k_c)O$ directly at this point:
\begin{equation}
O^{-1}H(\mathbf k_c)O=\begin{pmatrix} 2h_0 & 0 & 0 & 0\\ 0 & 0 & 0 &0 \\ 0 & 0 & 0 &0\\ 0 & 0 & 0 &-2h_0\end{pmatrix},
\end{equation}
where
\begin{equation}
O = \begin{pmatrix} \frac{h_0+\Omega}{\Delta} & \frac{\Omega}{\Delta} & \frac{h_0}{\Delta} & \frac{-h_0+\Omega}{\Delta}\\
\frac{h_0+\Omega}{\Delta} & -\frac{h_0}{\Delta} & -\frac{\Omega}{\Delta}& \frac{h_0-\Omega}{\Delta}\\
-1 & 0 & 1& 1\\ 1 & 1& 0 & 1\end{pmatrix}.
\end{equation}
For notational clarity, we apply a similarity transformation using the matrices $O$ and $O^{-1}$, rather than a unitary transformation with $O$ and $O^\dagger$. The difference is a simple normalization of the column vectors of $O$, and the underlying physics is not affected.

It is more instructive to look at the same similarity transformation applied to $\nabla_{\mathbf k}H(\mathbf k_c)$. We use the identities

\begin{eqnarray}
O^{-1}(\tau^z\otimes \mathbb I) O &=& \begin{pmatrix} \frac{\Omega}{h_0} & \frac{-h_0+\Omega}{2h_0} & \frac{h_0-\Omega}{2h_0} & 0\\-\frac{(h_0+\Omega)}{h_0} & 0 &\frac{\Omega}{h_0} & \frac{-h_0+\Omega}{h_0}\\ \frac{h_0+\Omega}{h_0} & \frac{\Omega}{h_0} & 0 &\frac{-h_0+\Omega}{h_0}\\ 0 & -\frac{(h_0+\Omega)}{2h_0} & -\frac{(h_0+\Omega)}{h_0} & -\frac{\Omega}{h_0}\end{pmatrix} \nonumber \\
O^{-1}(\mathbb I\otimes \sigma^z)O &=& \begin{pmatrix}0 & \frac{\Omega}{2h_0} & \frac{\Omega}{2h_0} &\frac{-h_0+\Omega}{h_0} \\ \frac{\Omega}{h_0} & -1 & -\frac{\Omega}{h_0} & -\frac{\Omega}{h_0} \\ \frac{\Omega}{h_0} & \frac{\Omega}{h_0} & 1 & \frac{\Omega}{h_0} \\ -\frac{(h_0+\Omega)}{h_0} & -\frac{\Omega}{2h_0} & \frac{\Omega}{2h_0} & 0\end{pmatrix}
\end{eqnarray}

The structure of the low energy surface states should follow from only the gapless bands. We therefore apply the similarity transformation $O^{-1}\nabla_{\mathbf k}H(\mathbf k_c)O$ and extract the second and the third (gapless) bands:
\begin{eqnarray}
O^{-1}(\tau^z\otimes \mathbb I) O & \to & \frac{\Omega}{h_0}\begin{pmatrix} 0 & 1\\ 1 & 0\end{pmatrix} = \frac{\Omega}{h_0}\bar{\sigma}^x,\\
O^{-1}(\mathbb I\otimes \sigma^z)O & \to & -\begin{pmatrix} 1 & 0 \\ 0 & -1\end{pmatrix} -i\frac{\Omega}{h_0}\begin{pmatrix} 0 & -i\\ i & 0\end{pmatrix} \nonumber,\\
& = & -\bar{\sigma}^z-i\frac{\Omega}{h_0}\bar{\sigma}^y.
\end{eqnarray}
Here we have defined new Pauli matrices $\bar{\sigma}^i$ that act only in the two-band low-energy subspace near the nodal points.
The low-energy subspace

\begin{equation}
\nabla_{\mathbf k}H(\mathbf k_c)\cdot\mathbf q\to q_z v_{\rm so}\Big(\bar{\sigma}^z+\frac{\Omega}{h_0}i\bar{\sigma}^y\Big)\pm\frac{\Omega k_x^{+}}{mh_0}q_x\bar{\sigma}^x. \label{eq:Heffkxp}
\end{equation}
Since $\{\bar{\sigma}^x, \bar{\sigma}^z+i \bar{\sigma}^y\Omega/h_0\}=0$, after defining the constant $v_1 = \Omega k_x^+/mh_0$, and the Pauli matrices $\bar{\sigma}^1 = \bar{\sigma}^x$ and $\bar{\sigma}^2 = \bar{\sigma}^z+i \bar{\sigma}^y\Omega/h_0$, we obtain Eq.~(7) in the main text:
\begin{equation}
\nabla_{\mathbf k} \bar{H}_{0}(\mathbf k_c)\cdot\mathbf{q} = \pm v_1 q_x \bar{\sigma}_1 + v_{\rm so} q_z  \bar{\sigma}_2
\end{equation}
This has the form of a Dirac Hamiltonian $q_{\mu}\gamma^{\mu}$ with $\{\gamma^i, \gamma^j\} = 2\delta^{ij}$.

The signature $\pm$ before $k_x^{+}$, corresponding to excitations near the points $\pm k^{+}_x$, defines distinct helicities of the excitations; thus, they are Weyl fermions. Moreover, if we regard $q_x$ as a parameter, i.e., applying a dimensional reduction, then the $\bar{\sigma}^1$ term is an effective ``{\it Dirac mass}.'' In this low energy effective approximation, the phase transition reflects the sign of $q_x$, which is associated with the sign of Dirac mass term. Breaking translational invariance in
the $\hat{z}$ direction by replacing $q_z\to -i\partial_z$, we can consider an interface at $z=0$ separating inequivalent phases $q_x>0$ and $q_x'<0.$ We have
\begin{align}
H_{\mathrm{eff}}= -i\partial_z\gamma^z+M(z)\gamma^x,
\end{align}
where $M$ takes the role of $v_1 q_x$, and is a positive constant for $z>0$ and a negative constant for $z<0$. The Jackiw and Rebbi or the Su-Schrieffer-Heeger~
\cite{PhysRevD.13.3398, PhysRevLett.42.1698} 
story is recovered and hence we establish the bulk-edge correspondence and confirm the existence of Majorana surface states at the $z=0$ interface by solving $H_{\mathrm{eff}}\psi(z) = 0.$

To be more precise, we should be able to distinguish the trivial and topological phases. If we look at $+k_x^+$ ($-k_x^+$), the Dirac mass becomes negative when $q_x<0$ ($q_x>0$). This is consistent with other analyses.
On the other hand, if we expand around $\pm k_x^-$, we can again locally diagonalize $H({\bf k}_c)$ with a different similarity matrix
\begin{equation}
 O'=\begin{pmatrix} \frac{h_0-\Omega}{\Delta} & -\frac{\Omega}{\Delta} & \frac{h_0}{\Delta} & -\frac{(h_0+\Omega)}{\Delta}\\
\frac{h_0-\Omega}{\Delta} & -\frac{h_0}{\Delta} & \frac{\Omega}{\Delta}& \frac{h_0+\Omega}{\Delta}\\
-1 & 0 & 1& 1\\ 1 & 1& 0 & 1\end{pmatrix},
\end{equation}
which produces the similar matrices
\begin{eqnarray}
{O'}^{-1}(\tau^z\otimes \mathbb I) O' & = & \begin{pmatrix} -\frac{\Omega}{h_0} & -\frac{(h_0+\Omega)}{2h_0} & \frac{h_0+\Omega}{2h_0} & 0\\ \frac{-h_0+\Omega}{h_0} & 0 &-\frac{\Omega}{h_0} & -\frac{(h_0+\Omega)}{h_0}\\ \frac{h_0-\Omega}{h_0} & -\frac{\Omega}{h_0} & 0 & -\frac{(h_0+\Omega)}{h_0}\\ 0 & \frac{-h_0+\Omega}{2h_0} & \frac{-h_0+\Omega}{h_0} & \frac{\Omega}{h_0}\end{pmatrix}, \nonumber\\
{O'}^{-1}(\mathbb I\otimes \sigma^z)O' & = & \begin{pmatrix}0 & -\frac{\Omega}{2h_0} & -\frac{\Omega}{2h_0} &-\frac{(h_0+\Omega)}{h_0} \\ -\frac{\Omega}{h_0} & -1 & \frac{\Omega}{h_0} & \frac{\Omega}{h_0} \\ -\frac{\Omega}{h_0} & -\frac{\Omega}{h_0} & 1 & -\frac{\Omega}{h_0} \\ \frac{-h_0+\Omega}{h_0} & \frac{\Omega}{2h_0} & -\frac{\Omega}{2h_0} & 0\end{pmatrix}.
\end{eqnarray}
Finally, we again project into the gapless subspace, and arrive at the low-energy effective Hamiltonian

\begin{equation}
\nabla_{\mathbf k}H(\mathbf k_c)\cdot\mathbf q\to q_z v_{\rm so} \left(\bar{\sigma}^z-\frac{\Omega}{h_0}i\bar{\sigma}^y \right)-(\pm)\frac{\Omega k_x^{-}}{mh_0}q_x\bar{\sigma}^x.
\end{equation}
This is identical, up to signs and a replacement $k_x^+ \rightarrow k_x^-$, to the effective Hamiltonian in Eq.~\ref{eq:Heffkxp} found by expanding around the second set of Dirac points.
The additional $(-1)$ appearing before $\sigma^x$ is necessary to make the criterion for topological phases consistent with the above arguments if we define $\gamma^x$ consistently.

Superficially, this approximation seems to depend on the ordering of eigenfunctions corresponding to the two gapless bands when constructing $O$ and $O'$. However, one can show straightforwardly that such an ambiguity does not change the relevant part of $O^{-1}(\tau^z\otimes\mathbb I)O$ and ${O'}^{-1}(\tau^z\otimes\mathbb I)O'$.
Our results are thus robust.

Finally, note that the topological protection is a property of the full four-band BdG Hamiltonian, and is based upon the existence of chiral symmetry. This argument does not rely on the structure of the low-energy effective expansion. Provided chiral symmetry remains, no perturbations can emerge that will result in a mass term.

\bibliography{Review}

\begin{thebibliography}{55}%
\makeatletter
\providecommand \@ifxundefined [1]{%
 \@ifx{#1\undefined}
}%
\providecommand \@ifnum [1]{%
 \ifnum #1\expandafter \@firstoftwo
 \else \expandafter \@secondoftwo
 \fi
}%
\providecommand \@ifx [1]{%
 \ifx #1\expandafter \@firstoftwo
 \else \expandafter \@secondoftwo
 \fi
}%
\providecommand \natexlab [1]{#1}%
\providecommand \enquote  [1]{``#1''}%
\providecommand \bibnamefont  [1]{#1}%
\providecommand \bibfnamefont [1]{#1}%
\providecommand \citenamefont [1]{#1}%
\providecommand \href@noop [0]{\@secondoftwo}%
\providecommand \href [0]{\begingroup \@sanitize@url \@href}%
\providecommand \@href[1]{\@@startlink{#1}\@@href}%
\providecommand \@@href[1]{\endgroup#1\@@endlink}%
\providecommand \@sanitize@url [0]{\catcode `\\12\catcode `\$12\catcode
  `\&12\catcode `\#12\catcode `\^12\catcode `\_12\catcode `\%12\relax}%
\providecommand \@@startlink[1]{}%
\providecommand \@@endlink[0]{}%
\providecommand \url  [0]{\begingroup\@sanitize@url \@url }%
\providecommand \@url [1]{\endgroup\@href {#1}{\urlprefix }}%
\providecommand \urlprefix  [0]{URL }%
\providecommand \Eprint [0]{\href }%
\providecommand \doibase [0]{http://dx.doi.org/}%
\providecommand \selectlanguage [0]{\@gobble}%
\providecommand \bibinfo  [0]{\@secondoftwo}%
\providecommand \bibfield  [0]{\@secondoftwo}%
\providecommand \translation [1]{[#1]}%
\providecommand \BibitemOpen [0]{}%
\providecommand \bibitemStop [0]{}%
\providecommand \bibitemNoStop [0]{.\EOS\space}%
\providecommand \EOS [0]{\spacefactor3000\relax}%
\providecommand \BibitemShut  [1]{\csname bibitem#1\endcsname}%
\let\auto@bib@innerbib\@empty
\bibitem [{\citenamefont {Jackiw}\ and\ \citenamefont
  {Rebbi}(1976)}]{PhysRevD.13.3398}%
  \BibitemOpen
  \bibfield  {author} {\bibinfo {author} {\bibfnamefont {R.}~\bibnamefont
  {Jackiw}}\ and\ \bibinfo {author} {\bibfnamefont {C.}~\bibnamefont {Rebbi}},\
  }\href {\doibase 10.1103/PhysRevD.13.3398} {\bibfield  {journal} {\bibinfo
  {journal} {Phys. Rev. D}\ }\textbf {\bibinfo {volume} {13}},\ \bibinfo
  {pages} {3398} (\bibinfo {year} {1976})}\BibitemShut {NoStop}%
\bibitem [{\citenamefont {Xia-Ji}\ \emph {et~al.}(2015)\citenamefont {Xia-Ji},
  \citenamefont {Hui},\ and\ \citenamefont {Han}}]{1674-1056-24-5-050502}%
  \BibitemOpen
  \bibfield  {author} {\bibinfo {author} {\bibfnamefont {L.}~\bibnamefont
  {Xia-Ji}}, \bibinfo {author} {\bibfnamefont {H.}~\bibnamefont {Hui}}, \ and\
  \bibinfo {author} {\bibfnamefont {P.}~\bibnamefont {Han}},\ }\href
  {http://stacks.iop.org/1674-1056/24/i=5/a=050502} {\bibfield  {journal}
  {\bibinfo  {journal} {Chinese Physics B}\ }\textbf {\bibinfo {volume} {24}},\
  \bibinfo {pages} {050502} (\bibinfo {year} {2015})}\BibitemShut {NoStop}%
\bibitem [{\citenamefont {Xu}\ and\ \citenamefont
  {Zhang}(2016)}]{PhysRevA.93.063606}%
  \BibitemOpen
  \bibfield  {author} {\bibinfo {author} {\bibfnamefont {Y.}~\bibnamefont
  {Xu}}\ and\ \bibinfo {author} {\bibfnamefont {C.}~\bibnamefont {Zhang}},\
  }\href {\doibase 10.1103/PhysRevA.93.063606} {\bibfield  {journal} {\bibinfo
  {journal} {Phys. Rev. A}\ }\textbf {\bibinfo {volume} {93}},\ \bibinfo
  {pages} {063606} (\bibinfo {year} {2016})}\BibitemShut {NoStop}%
\bibitem [{\citenamefont {Seo}\ \emph {et~al.}(2012)\citenamefont {Seo},
  \citenamefont {Han},\ and\ \citenamefont {S\'a~de
  Melo}}]{PhysRevLett.109.105303}%
  \BibitemOpen
  \bibfield  {author} {\bibinfo {author} {\bibfnamefont {K.}~\bibnamefont
  {Seo}}, \bibinfo {author} {\bibfnamefont {L.}~\bibnamefont {Han}}, \ and\
  \bibinfo {author} {\bibfnamefont {C.~A.~R.}\ \bibnamefont {S\'a~de Melo}},\
  }\href {\doibase 10.1103/PhysRevLett.109.105303} {\bibfield  {journal}
  {\bibinfo  {journal} {Phys. Rev. Lett.}\ }\textbf {\bibinfo {volume} {109}},\
  \bibinfo {pages} {105303} (\bibinfo {year} {2012})}\BibitemShut {NoStop}%
\bibitem [{\citenamefont {Fu}\ and\ \citenamefont
  {Kane}(2008{\natexlab{a}})}]{fu}%
  \BibitemOpen
  \bibfield  {author} {\bibinfo {author} {\bibfnamefont {L.}~\bibnamefont
  {Fu}}\ and\ \bibinfo {author} {\bibfnamefont {C.~L.}\ \bibnamefont {Kane}},\
  }\href {\doibase 10.1103/PhysRevLett.100.096407} {\bibfield  {journal}
  {\bibinfo  {journal} {Phys. Rev. Lett.}\ }\textbf {\bibinfo {volume} {100}},\
  \bibinfo {pages} {096407} (\bibinfo {year} {2008}{\natexlab{a}})}\BibitemShut
  {NoStop}%
\bibitem [{\citenamefont {Sau}\ \emph {et~al.}(2010{\natexlab{a}})\citenamefont
  {Sau}, \citenamefont {Lutchyn}, \citenamefont {Tewari},\ and\ \citenamefont
  {Das~Sarma}}]{sau}%
  \BibitemOpen
  \bibfield  {author} {\bibinfo {author} {\bibfnamefont {J.~D.}\ \bibnamefont
  {Sau}}, \bibinfo {author} {\bibfnamefont {R.~M.}\ \bibnamefont {Lutchyn}},
  \bibinfo {author} {\bibfnamefont {S.}~\bibnamefont {Tewari}}, \ and\ \bibinfo
  {author} {\bibfnamefont {S.}~\bibnamefont {Das~Sarma}},\ }\href {\doibase
  10.1103/PhysRevLett.104.040502} {\bibfield  {journal} {\bibinfo  {journal}
  {Phys. Rev. Lett.}\ }\textbf {\bibinfo {volume} {104}},\ \bibinfo {pages}
  {040502} (\bibinfo {year} {2010}{\natexlab{a}})}\BibitemShut {NoStop}%
\bibitem [{\citenamefont {Fu}\ and\ \citenamefont
  {Kane}(2008{\natexlab{b}})}]{PhysRevLett.100.096407}%
  \BibitemOpen
  \bibfield  {author} {\bibinfo {author} {\bibfnamefont {L.}~\bibnamefont
  {Fu}}\ and\ \bibinfo {author} {\bibfnamefont {C.~L.}\ \bibnamefont {Kane}},\
  }\href {\doibase 10.1103/PhysRevLett.100.096407} {\bibfield  {journal}
  {\bibinfo  {journal} {Phys. Rev. Lett.}\ }\textbf {\bibinfo {volume} {100}},\
  \bibinfo {pages} {096407} (\bibinfo {year} {2008}{\natexlab{b}})}\BibitemShut
  {NoStop}%
\bibitem [{\citenamefont {Sau}\ \emph {et~al.}(2010{\natexlab{b}})\citenamefont
  {Sau}, \citenamefont {Lutchyn}, \citenamefont {Tewari},\ and\ \citenamefont
  {Das~Sarma}}]{PhysRevLett.104.040502}%
  \BibitemOpen
  \bibfield  {author} {\bibinfo {author} {\bibfnamefont {J.~D.}\ \bibnamefont
  {Sau}}, \bibinfo {author} {\bibfnamefont {R.~M.}\ \bibnamefont {Lutchyn}},
  \bibinfo {author} {\bibfnamefont {S.}~\bibnamefont {Tewari}}, \ and\ \bibinfo
  {author} {\bibfnamefont {S.}~\bibnamefont {Das~Sarma}},\ }\href {\doibase
  10.1103/PhysRevLett.104.040502} {\bibfield  {journal} {\bibinfo  {journal}
  {Phys. Rev. Lett.}\ }\textbf {\bibinfo {volume} {104}},\ \bibinfo {pages}
  {040502} (\bibinfo {year} {2010}{\natexlab{b}})}\BibitemShut {NoStop}%
\bibitem [{\citenamefont {Linder}\ \emph {et~al.}(2010)\citenamefont {Linder},
  \citenamefont {Tanaka}, \citenamefont {Yokoyama}, \citenamefont {Sudb\o{}},\
  and\ \citenamefont {Nagaosa}}]{PhysRevLett.104.067001}%
  \BibitemOpen
  \bibfield  {author} {\bibinfo {author} {\bibfnamefont {J.}~\bibnamefont
  {Linder}}, \bibinfo {author} {\bibfnamefont {Y.}~\bibnamefont {Tanaka}},
  \bibinfo {author} {\bibfnamefont {T.}~\bibnamefont {Yokoyama}}, \bibinfo
  {author} {\bibfnamefont {A.}~\bibnamefont {Sudb\o{}}}, \ and\ \bibinfo
  {author} {\bibfnamefont {N.}~\bibnamefont {Nagaosa}},\ }\href {\doibase
  10.1103/PhysRevLett.104.067001} {\bibfield  {journal} {\bibinfo  {journal}
  {Phys. Rev. Lett.}\ }\textbf {\bibinfo {volume} {104}},\ \bibinfo {pages}
  {067001} (\bibinfo {year} {2010})}\BibitemShut {NoStop}%
\bibitem [{\citenamefont {Lutchyn}\ \emph {et~al.}(2010)\citenamefont
  {Lutchyn}, \citenamefont {Sau},\ and\ \citenamefont
  {Das~Sarma}}]{PhysRevLett.105.077001}%
  \BibitemOpen
  \bibfield  {author} {\bibinfo {author} {\bibfnamefont {R.~M.}\ \bibnamefont
  {Lutchyn}}, \bibinfo {author} {\bibfnamefont {J.~D.}\ \bibnamefont {Sau}}, \
  and\ \bibinfo {author} {\bibfnamefont {S.}~\bibnamefont {Das~Sarma}},\ }\href
  {\doibase 10.1103/PhysRevLett.105.077001} {\bibfield  {journal} {\bibinfo
  {journal} {Phys. Rev. Lett.}\ }\textbf {\bibinfo {volume} {105}},\ \bibinfo
  {pages} {077001} (\bibinfo {year} {2010})}\BibitemShut {NoStop}%
\bibitem [{\citenamefont {Oreg}\ \emph {et~al.}(2010)\citenamefont {Oreg},
  \citenamefont {Refael},\ and\ \citenamefont {von
  Oppen}}]{PhysRevLett.105.177002}%
  \BibitemOpen
  \bibfield  {author} {\bibinfo {author} {\bibfnamefont {Y.}~\bibnamefont
  {Oreg}}, \bibinfo {author} {\bibfnamefont {G.}~\bibnamefont {Refael}}, \ and\
  \bibinfo {author} {\bibfnamefont {F.}~\bibnamefont {von Oppen}},\ }\href
  {\doibase 10.1103/PhysRevLett.105.177002} {\bibfield  {journal} {\bibinfo
  {journal} {Phys. Rev. Lett.}\ }\textbf {\bibinfo {volume} {105}},\ \bibinfo
  {pages} {177002} (\bibinfo {year} {2010})}\BibitemShut {NoStop}%
\bibitem [{\citenamefont {Wu}\ \emph {et~al.}(2012)\citenamefont {Wu},
  \citenamefont {Valls},\ and\ \citenamefont {Halterman}}]{wu1}%
  \BibitemOpen
  \bibfield  {author} {\bibinfo {author} {\bibfnamefont {C.-T.}\ \bibnamefont
  {Wu}}, \bibinfo {author} {\bibfnamefont {O.~T.}\ \bibnamefont {Valls}}, \
  and\ \bibinfo {author} {\bibfnamefont {K.}~\bibnamefont {Halterman}},\ }\href
  {\doibase 10.1103/PhysRevB.86.184517} {\bibfield  {journal} {\bibinfo
  {journal} {Phys. Rev. B}\ }\textbf {\bibinfo {volume} {86}},\ \bibinfo
  {pages} {184517} (\bibinfo {year} {2012})}\BibitemShut {NoStop}%
\bibitem [{\citenamefont {Robinson}\ \emph {et~al.}(2010)\citenamefont
  {Robinson}, \citenamefont {Witt},\ and\ \citenamefont {Blamire}}]{robinson}%
  \BibitemOpen
  \bibfield  {author} {\bibinfo {author} {\bibfnamefont {J.~W.~A.}\
  \bibnamefont {Robinson}}, \bibinfo {author} {\bibfnamefont {J.~D.~S.}\
  \bibnamefont {Witt}}, \ and\ \bibinfo {author} {\bibfnamefont {M.~G.}\
  \bibnamefont {Blamire}},\ }\href {\doibase 10.1126/science.1189246}
  {\bibfield  {journal} {\bibinfo  {journal} {Science}\ }\textbf {\bibinfo
  {volume} {329}},\ \bibinfo {pages} {59} (\bibinfo {year} {2010})}\BibitemShut
  {NoStop}%
\bibitem [{\citenamefont {Chiodi}\ \emph {et~al.}(2013)\citenamefont {Chiodi},
  \citenamefont {Witt}, \citenamefont {Smits}, \citenamefont {Qu},
  \citenamefont {Halasz}, \citenamefont {Wu}, \citenamefont {Valls},
  \citenamefont {Halterman}, \citenamefont {Robinson},\ and\ \citenamefont
  {Blamire}}]{chiodi}%
  \BibitemOpen
  \bibfield  {author} {\bibinfo {author} {\bibfnamefont {F.}~\bibnamefont
  {Chiodi}}, \bibinfo {author} {\bibfnamefont {J.~D.~S.}\ \bibnamefont {Witt}},
  \bibinfo {author} {\bibfnamefont {R.~G.~J.}\ \bibnamefont {Smits}}, \bibinfo
  {author} {\bibfnamefont {L.}~\bibnamefont {Qu}}, \bibinfo {author}
  {\bibfnamefont {G.~B.}\ \bibnamefont {Halasz}}, \bibinfo {author}
  {\bibfnamefont {C.-T.}\ \bibnamefont {Wu}}, \bibinfo {author} {\bibfnamefont
  {O.~T.}\ \bibnamefont {Valls}}, \bibinfo {author} {\bibfnamefont
  {K.}~\bibnamefont {Halterman}}, \bibinfo {author} {\bibfnamefont {J.~W.~A.}\
  \bibnamefont {Robinson}}, \ and\ \bibinfo {author} {\bibfnamefont {M.~G.}\
  \bibnamefont {Blamire}},\ }\href
  {http://stacks.iop.org/0295-5075/101/i=3/a=37002} {\bibfield  {journal}
  {\bibinfo  {journal} {EPL}\ }\textbf {\bibinfo {volume} {101}},\ \bibinfo
  {pages} {37002} (\bibinfo {year} {2013})}\BibitemShut {NoStop}%
\bibitem [{\citenamefont {{Liu}}\ \emph {et~al.}(2015)\citenamefont {{Liu}},
  \citenamefont {{Hu}},\ and\ \citenamefont {{Pu}}}]{2015ChPhB..24e0502L}%
  \BibitemOpen
  \bibfield  {author} {\bibinfo {author} {\bibfnamefont {X.-J.}\ \bibnamefont
  {{Liu}}}, \bibinfo {author} {\bibfnamefont {H.}~\bibnamefont {{Hu}}}, \ and\
  \bibinfo {author} {\bibfnamefont {H.}~\bibnamefont {{Pu}}},\ }\href {\doibase
  10.1088/1674-1056/24/5/050502} {\bibfield  {journal} {\bibinfo  {journal}
  {Chinese Physics B}\ }\textbf {\bibinfo {volume} {24}},\ \bibinfo {eid}
  {050502} (\bibinfo {year} {2015})},\ \Eprint {http://arxiv.org/abs/1411.2993}
  {arXiv:1411.2993 [cond-mat.quant-gas]} \BibitemShut {NoStop}%
\bibitem [{\citenamefont {Xu}\ \emph {et~al.}(2015)\citenamefont {Xu},
  \citenamefont {Zhang},\ and\ \citenamefont {Zhang}}]{PhysRevLett.115.265304}%
  \BibitemOpen
  \bibfield  {author} {\bibinfo {author} {\bibfnamefont {Y.}~\bibnamefont
  {Xu}}, \bibinfo {author} {\bibfnamefont {F.}~\bibnamefont {Zhang}}, \ and\
  \bibinfo {author} {\bibfnamefont {C.}~\bibnamefont {Zhang}},\ }\href
  {\doibase 10.1103/PhysRevLett.115.265304} {\bibfield  {journal} {\bibinfo
  {journal} {Phys. Rev. Lett.}\ }\textbf {\bibinfo {volume} {115}},\ \bibinfo
  {pages} {265304} (\bibinfo {year} {2015})}\BibitemShut {NoStop}%
\bibitem [{\citenamefont {Nadj-Perge}\ \emph {et~al.}(2014)\citenamefont
  {Nadj-Perge}, \citenamefont {Drozdov}, \citenamefont {Li}, \citenamefont
  {Chen}, \citenamefont {Sangjun}, \citenamefont {Seo}, \citenamefont
  {MacDonald}, \citenamefont {Bernevig},\ and\ \citenamefont {Yazdani}}]{Yaz1}%
  \BibitemOpen
  \bibfield  {author} {\bibinfo {author} {\bibfnamefont {S.}~\bibnamefont
  {Nadj-Perge}}, \bibinfo {author} {\bibfnamefont {I.~K.}\ \bibnamefont
  {Drozdov}}, \bibinfo {author} {\bibfnamefont {J.}~\bibnamefont {Li}},
  \bibinfo {author} {\bibfnamefont {H.}~\bibnamefont {Chen}}, \bibinfo {author}
  {\bibfnamefont {J.}~\bibnamefont {Sangjun}}, \bibinfo {author} {\bibfnamefont
  {J.}~\bibnamefont {Seo}}, \bibinfo {author} {\bibfnamefont {A.~H.}\
  \bibnamefont {MacDonald}}, \bibinfo {author} {\bibfnamefont {B.}~\bibnamefont
  {Bernevig}}, \ and\ \bibinfo {author} {\bibfnamefont {A.}~\bibnamefont
  {Yazdani}},\ }\href@noop {} {\bibfield  {journal} {\bibinfo  {journal}
  {Science}\ }\textbf {\bibinfo {volume} {346}},\ \bibinfo {pages} {602}
  (\bibinfo {year} {2014})}\BibitemShut {NoStop}%
\bibitem [{\citenamefont {Klinovaja}\ \emph {et~al.}(2013)\citenamefont
  {Klinovaja}, \citenamefont {Stano}, \citenamefont {Yazdani},\ and\
  \citenamefont {Loss}}]{Yaz2}%
  \BibitemOpen
  \bibfield  {author} {\bibinfo {author} {\bibfnamefont {J.}~\bibnamefont
  {Klinovaja}}, \bibinfo {author} {\bibfnamefont {P.}~\bibnamefont {Stano}},
  \bibinfo {author} {\bibfnamefont {A.}~\bibnamefont {Yazdani}}, \ and\
  \bibinfo {author} {\bibfnamefont {D.}~\bibnamefont {Loss}},\ }\href {\doibase
  10.1103/PhysRevLett.111.186805} {\bibfield  {journal} {\bibinfo  {journal}
  {Phys. Rev. Lett.}\ }\textbf {\bibinfo {volume} {111}},\ \bibinfo {pages}
  {186805} (\bibinfo {year} {2013})}\BibitemShut {NoStop}%
\bibitem [{\citenamefont {Nadj-Perge}\ \emph {et~al.}(2013)\citenamefont
  {Nadj-Perge}, \citenamefont {Drozdov}, \citenamefont {Bernevig},\ and\
  \citenamefont {Yazdani}}]{Yaz3}%
  \BibitemOpen
  \bibfield  {author} {\bibinfo {author} {\bibfnamefont {S.}~\bibnamefont
  {Nadj-Perge}}, \bibinfo {author} {\bibfnamefont {I.~K.}\ \bibnamefont
  {Drozdov}}, \bibinfo {author} {\bibfnamefont {B.~A.}\ \bibnamefont
  {Bernevig}}, \ and\ \bibinfo {author} {\bibfnamefont {A.}~\bibnamefont
  {Yazdani}},\ }\href {\doibase 10.1103/PhysRevB.88.020407} {\bibfield
  {journal} {\bibinfo  {journal} {Phys. Rev. B}\ }\textbf {\bibinfo {volume}
  {88}},\ \bibinfo {pages} {020407} (\bibinfo {year} {2013})}\BibitemShut
  {NoStop}%
\bibitem [{\citenamefont {Li}\ \emph {et~al.}(2014)\citenamefont {Li},
  \citenamefont {Chen}, \citenamefont {Drozdov}, \citenamefont {Yazdani},
  \citenamefont {Bernevig},\ and\ \citenamefont {MacDonald}}]{Yaz4}%
  \BibitemOpen
  \bibfield  {author} {\bibinfo {author} {\bibfnamefont {J.}~\bibnamefont
  {Li}}, \bibinfo {author} {\bibfnamefont {H.}~\bibnamefont {Chen}}, \bibinfo
  {author} {\bibfnamefont {I.~K.}\ \bibnamefont {Drozdov}}, \bibinfo {author}
  {\bibfnamefont {A.}~\bibnamefont {Yazdani}}, \bibinfo {author} {\bibfnamefont
  {B.~A.}\ \bibnamefont {Bernevig}}, \ and\ \bibinfo {author} {\bibfnamefont
  {A.~H.}\ \bibnamefont {MacDonald}},\ }\href {\doibase
  10.1103/PhysRevB.90.235433} {\bibfield  {journal} {\bibinfo  {journal} {Phys.
  Rev. B}\ }\textbf {\bibinfo {volume} {90}},\ \bibinfo {pages} {235433}
  (\bibinfo {year} {2014})}\BibitemShut {NoStop}%
\bibitem [{\citenamefont {Li}\ \emph {et~al.}(2016)\citenamefont {Li},
  \citenamefont {Neupert}, \citenamefont {Wang}, \citenamefont {MacDonald},
  \citenamefont {Yazdani},\ and\ \citenamefont {Bernevig}}]{Yaz5}%
  \BibitemOpen
  \bibfield  {author} {\bibinfo {author} {\bibfnamefont {J.}~\bibnamefont
  {Li}}, \bibinfo {author} {\bibfnamefont {T.}~\bibnamefont {Neupert}},
  \bibinfo {author} {\bibfnamefont {Z.}~\bibnamefont {Wang}}, \bibinfo {author}
  {\bibfnamefont {A.~H.}\ \bibnamefont {MacDonald}}, \bibinfo {author}
  {\bibfnamefont {Y.}~\bibnamefont {Yazdani}}, \ and\ \bibinfo {author}
  {\bibfnamefont {A.}~\bibnamefont {Bernevig}},\ }\href@noop {} {\bibfield
  {journal} {\bibinfo  {journal} {Nature Communications}\ }\textbf {\bibinfo
  {volume} {7}},\ \bibinfo {pages} {12297} (\bibinfo {year}
  {2016})}\BibitemShut {NoStop}%
\bibitem [{\citenamefont {Sosnin}\ \emph {et~al.}(2006)\citenamefont {Sosnin},
  \citenamefont {Cho}, \citenamefont {Petrashov},\ and\ \citenamefont
  {Volkov}}]{Volkov3}%
  \BibitemOpen
  \bibfield  {author} {\bibinfo {author} {\bibfnamefont {I.}~\bibnamefont
  {Sosnin}}, \bibinfo {author} {\bibfnamefont {H.}~\bibnamefont {Cho}},
  \bibinfo {author} {\bibfnamefont {V.~T.}\ \bibnamefont {Petrashov}}, \ and\
  \bibinfo {author} {\bibfnamefont {A.~F.}\ \bibnamefont {Volkov}},\ }\href
  {\doibase 10.1103/PhysRevLett.96.157002} {\bibfield  {journal} {\bibinfo
  {journal} {Phys. Rev. Lett.}\ }\textbf {\bibinfo {volume} {96}},\ \bibinfo
  {pages} {157002} (\bibinfo {year} {2006})}\BibitemShut {NoStop}%
\bibitem [{\citenamefont {Halterman}\ \emph {et~al.}(2007)\citenamefont
  {Halterman}, \citenamefont {Barsic},\ and\ \citenamefont
  {Valls}}]{halterman1}%
  \BibitemOpen
  \bibfield  {author} {\bibinfo {author} {\bibfnamefont {K.}~\bibnamefont
  {Halterman}}, \bibinfo {author} {\bibfnamefont {P.~H.}\ \bibnamefont
  {Barsic}}, \ and\ \bibinfo {author} {\bibfnamefont {O.~T.}\ \bibnamefont
  {Valls}},\ }\href {\doibase 10.1103/PhysRevLett.99.127002} {\bibfield
  {journal} {\bibinfo  {journal} {Phys. Rev. Lett.}\ }\textbf {\bibinfo
  {volume} {99}},\ \bibinfo {pages} {127002} (\bibinfo {year}
  {2007})}\BibitemShut {NoStop}%
\bibitem [{\citenamefont {Halterman}\ \emph {et~al.}(2008)\citenamefont
  {Halterman}, \citenamefont {Valls},\ and\ \citenamefont
  {Barsic}}]{halterman2}%
  \BibitemOpen
  \bibfield  {author} {\bibinfo {author} {\bibfnamefont {K.}~\bibnamefont
  {Halterman}}, \bibinfo {author} {\bibfnamefont {O.~T.}\ \bibnamefont
  {Valls}}, \ and\ \bibinfo {author} {\bibfnamefont {P.~H.}\ \bibnamefont
  {Barsic}},\ }\href {\doibase 10.1103/PhysRevB.77.174511} {\bibfield
  {journal} {\bibinfo  {journal} {Phys. Rev. B}\ }\textbf {\bibinfo {volume}
  {77}},\ \bibinfo {pages} {174511} (\bibinfo {year} {2008})}\BibitemShut
  {NoStop}%
\bibitem [{\citenamefont {Bergeret}\ \emph {et~al.}(2001)\citenamefont
  {Bergeret}, \citenamefont {Volkov},\ and\ \citenamefont {Efetov}}]{Volkov1}%
  \BibitemOpen
  \bibfield  {author} {\bibinfo {author} {\bibfnamefont {F.~S.}\ \bibnamefont
  {Bergeret}}, \bibinfo {author} {\bibfnamefont {A.~F.}\ \bibnamefont
  {Volkov}}, \ and\ \bibinfo {author} {\bibfnamefont {K.~B.}\ \bibnamefont
  {Efetov}},\ }\href {\doibase 10.1103/PhysRevLett.86.4096} {\bibfield
  {journal} {\bibinfo  {journal} {Phys. Rev. Lett.}\ }\textbf {\bibinfo
  {volume} {86}},\ \bibinfo {pages} {4096} (\bibinfo {year}
  {2001})}\BibitemShut {NoStop}%
\bibitem [{\citenamefont {Demler}\ \emph {et~al.}(1997)\citenamefont {Demler},
  \citenamefont {Arnold},\ and\ \citenamefont {Beasley}}]{demler}%
  \BibitemOpen
  \bibfield  {author} {\bibinfo {author} {\bibfnamefont {E.~A.}\ \bibnamefont
  {Demler}}, \bibinfo {author} {\bibfnamefont {G.~B.}\ \bibnamefont {Arnold}},
  \ and\ \bibinfo {author} {\bibfnamefont {M.~R.}\ \bibnamefont {Beasley}},\
  }\href {\doibase 10.1103/PhysRevB.55.15174} {\bibfield  {journal} {\bibinfo
  {journal} {Phys. Rev. B}\ }\textbf {\bibinfo {volume} {55}},\ \bibinfo
  {pages} {15174} (\bibinfo {year} {1997})}\BibitemShut {NoStop}%
\bibitem [{\citenamefont {Halterman}\ and\ \citenamefont
  {Valls}(2001)}]{halterman3}%
  \BibitemOpen
  \bibfield  {author} {\bibinfo {author} {\bibfnamefont {K.}~\bibnamefont
  {Halterman}}\ and\ \bibinfo {author} {\bibfnamefont {O.~T.}\ \bibnamefont
  {Valls}},\ }\href {\doibase 10.1103/PhysRevB.65.014509} {\bibfield  {journal}
  {\bibinfo  {journal} {Phys. Rev. B}\ }\textbf {\bibinfo {volume} {65}},\
  \bibinfo {pages} {014509} (\bibinfo {year} {2001})}\BibitemShut {NoStop}%
\bibitem [{\citenamefont {Buzdin}(2005)}]{Buzdin}%
  \BibitemOpen
  \bibfield  {author} {\bibinfo {author} {\bibfnamefont {A.~I.}\ \bibnamefont
  {Buzdin}},\ }\href {\doibase 10.1103/RevModPhys.77.935} {\bibfield  {journal}
  {\bibinfo  {journal} {Rev. Mod. Phys.}\ }\textbf {\bibinfo {volume} {77}},\
  \bibinfo {pages} {935} (\bibinfo {year} {2005})}\BibitemShut {NoStop}%
\bibitem [{\citenamefont {Bergeret}\ \emph {et~al.}(2005)\citenamefont
  {Bergeret}, \citenamefont {Volkov},\ and\ \citenamefont {Efetov}}]{bergeret}%
  \BibitemOpen
  \bibfield  {author} {\bibinfo {author} {\bibfnamefont {F.~S.}\ \bibnamefont
  {Bergeret}}, \bibinfo {author} {\bibfnamefont {A.~F.}\ \bibnamefont
  {Volkov}}, \ and\ \bibinfo {author} {\bibfnamefont {K.~B.}\ \bibnamefont
  {Efetov}},\ }\href {\doibase 10.1103/RevModPhys.77.1321} {\bibfield
  {journal} {\bibinfo  {journal} {Rev. Mod. Phys.}\ }\textbf {\bibinfo {volume}
  {77}},\ \bibinfo {pages} {1321} (\bibinfo {year} {2005})}\BibitemShut
  {NoStop}%
\bibitem [{\citenamefont {Volkov}\ \emph {et~al.}(2006)\citenamefont {Volkov},
  \citenamefont {Anishchanka},\ and\ \citenamefont {Efetov}}]{Volkov2}%
  \BibitemOpen
  \bibfield  {author} {\bibinfo {author} {\bibfnamefont {A.~F.}\ \bibnamefont
  {Volkov}}, \bibinfo {author} {\bibfnamefont {A.}~\bibnamefont {Anishchanka}},
  \ and\ \bibinfo {author} {\bibfnamefont {K.~B.}\ \bibnamefont {Efetov}},\
  }\href {\doibase 10.1103/PhysRevB.73.104412} {\bibfield  {journal} {\bibinfo
  {journal} {Phys. Rev. B}\ }\textbf {\bibinfo {volume} {73}},\ \bibinfo
  {pages} {104412} (\bibinfo {year} {2006})}\BibitemShut {NoStop}%
\bibitem [{\citenamefont {Lababidi}\ and\ \citenamefont
  {Zhao}(2011)}]{Lababidi}%
  \BibitemOpen
  \bibfield  {author} {\bibinfo {author} {\bibfnamefont {M.}~\bibnamefont
  {Lababidi}}\ and\ \bibinfo {author} {\bibfnamefont {E.}~\bibnamefont
  {Zhao}},\ }\href {\doibase 10.1103/PhysRevB.83.184511} {\bibfield  {journal}
  {\bibinfo  {journal} {Phys. Rev. B}\ }\textbf {\bibinfo {volume} {83}},\
  \bibinfo {pages} {184511} (\bibinfo {year} {2011})}\BibitemShut {NoStop}%
\bibitem [{\citenamefont {Chiu}\ \emph {et~al.}(2016)\citenamefont {Chiu},
  \citenamefont {Cole},\ and\ \citenamefont {Das~Sarma}}]{dasSarmanew}%
  \BibitemOpen
  \bibfield  {author} {\bibinfo {author} {\bibfnamefont {C.-K.}\ \bibnamefont
  {Chiu}}, \bibinfo {author} {\bibfnamefont {W.~S.}\ \bibnamefont {Cole}}, \
  and\ \bibinfo {author} {\bibfnamefont {S.}~\bibnamefont {Das~Sarma}},\ }\href
  {\doibase 10.1103/PhysRevB.94.125304} {\bibfield  {journal} {\bibinfo
  {journal} {Phys. Rev. B}\ }\textbf {\bibinfo {volume} {94}},\ \bibinfo
  {pages} {125304} (\bibinfo {year} {2016})}\BibitemShut {NoStop}%
\bibitem [{\citenamefont {Linder}\ and\ \citenamefont
  {Robinson}(2015)}]{linder}%
  \BibitemOpen
  \bibfield  {author} {\bibinfo {author} {\bibfnamefont {J.}~\bibnamefont
  {Linder}}\ and\ \bibinfo {author} {\bibfnamefont {J.~W.~A.}\ \bibnamefont
  {Robinson}},\ }\href {\doibase 10.1038/nphys3242} {\bibfield  {journal}
  {\bibinfo  {journal} {Nature Physics}\ }\textbf {\bibinfo {volume} {11}},\
  \bibinfo {pages} {307} (\bibinfo {year} {2015})}\BibitemShut {NoStop}%
\bibitem [{\citenamefont {de~V.~du Plessis}\ \emph {et~al.}(1983)\citenamefont
  {de~V.~du Plessis}, \citenamefont {van Doorn},\ and\ \citenamefont {van
  Delden}}]{tbdy}%
  \BibitemOpen
  \bibfield  {author} {\bibinfo {author} {\bibfnamefont {P.}~\bibnamefont
  {de~V.~du Plessis}}, \bibinfo {author} {\bibfnamefont {C.~F.}\ \bibnamefont
  {van Doorn}}, \ and\ \bibinfo {author} {\bibfnamefont {D.~C.}\ \bibnamefont
  {van Delden}},\ }\href {\doibase doi:10.1016/0304-8853(83)90014-8} {\bibfield
   {journal} {\bibinfo  {journal} {J. Magn. Magn. Mater.}\ }\textbf {\bibinfo
  {volume} {40}},\ \bibinfo {pages} {91} (\bibinfo {year} {1983})}\BibitemShut
  {NoStop}%
\bibitem [{\citenamefont {Shirane}\ \emph {et~al.}(1983)\citenamefont
  {Shirane}, \citenamefont {Cowley}, \citenamefont {Majkrzak}, \citenamefont
  {Sokoloff}, \citenamefont {Pagonis}, \citenamefont {Perry},\ and\
  \citenamefont {Ishikawa}}]{mnsi}%
  \BibitemOpen
  \bibfield  {author} {\bibinfo {author} {\bibfnamefont {G.}~\bibnamefont
  {Shirane}}, \bibinfo {author} {\bibfnamefont {R.}~\bibnamefont {Cowley}},
  \bibinfo {author} {\bibfnamefont {C.}~\bibnamefont {Majkrzak}}, \bibinfo
  {author} {\bibfnamefont {J.~B.}\ \bibnamefont {Sokoloff}}, \bibinfo {author}
  {\bibfnamefont {B.}~\bibnamefont {Pagonis}}, \bibinfo {author} {\bibfnamefont
  {C.~H.}\ \bibnamefont {Perry}}, \ and\ \bibinfo {author} {\bibfnamefont
  {Y.}~\bibnamefont {Ishikawa}},\ }\href {\doibase 10.1103/PhysRevB.28.6251}
  {\bibfield  {journal} {\bibinfo  {journal} {Phys. Rev. B}\ }\textbf {\bibinfo
  {volume} {28}},\ \bibinfo {pages} {6251} (\bibinfo {year}
  {1983})}\BibitemShut {NoStop}%
\bibitem [{\citenamefont {Lin}\ \emph {et~al.}(2011)\citenamefont {Lin},
  \citenamefont {Jimenez-Garcia},\ and\ \citenamefont
  {Spielman}}]{spielmansoc}%
  \BibitemOpen
  \bibfield  {author} {\bibinfo {author} {\bibfnamefont {Y.~J.}\ \bibnamefont
  {Lin}}, \bibinfo {author} {\bibfnamefont {K.}~\bibnamefont {Jimenez-Garcia}},
  \ and\ \bibinfo {author} {\bibfnamefont {I.~B.}\ \bibnamefont {Spielman}},\
  }\href {http://dx.doi.org/10.1038/nature09887} {\bibfield  {journal}
  {\bibinfo  {journal} {Nature}\ }\textbf {\bibinfo {volume} {471}},\ \bibinfo
  {pages} {83} (\bibinfo {year} {2011})}\BibitemShut {NoStop}%
\bibitem [{\citenamefont {Martin}\ and\ \citenamefont
  {Morpurgo}(2012)}]{martin}%
  \BibitemOpen
  \bibfield  {author} {\bibinfo {author} {\bibfnamefont {I.}~\bibnamefont
  {Martin}}\ and\ \bibinfo {author} {\bibfnamefont {A.~F.}\ \bibnamefont
  {Morpurgo}},\ }\href {\doibase 10.1103/PhysRevB.85.144505} {\bibfield
  {journal} {\bibinfo  {journal} {Phys. Rev. B}\ }\textbf {\bibinfo {volume}
  {85}},\ \bibinfo {pages} {144505} (\bibinfo {year} {2012})}\BibitemShut
  {NoStop}%
\bibitem [{\citenamefont {Braunecker}\ \emph {et~al.}(2010)\citenamefont
  {Braunecker}, \citenamefont {Japaridze}, \citenamefont {Klinovaja},\ and\
  \citenamefont {Loss}}]{JelenaK}%
  \BibitemOpen
  \bibfield  {author} {\bibinfo {author} {\bibfnamefont {B.}~\bibnamefont
  {Braunecker}}, \bibinfo {author} {\bibfnamefont {G.~I.}\ \bibnamefont
  {Japaridze}}, \bibinfo {author} {\bibfnamefont {J.}~\bibnamefont
  {Klinovaja}}, \ and\ \bibinfo {author} {\bibfnamefont {D.}~\bibnamefont
  {Loss}},\ }\href {\doibase 10.1103/PhysRevB.82.045127} {\bibfield  {journal}
  {\bibinfo  {journal} {Phys. Rev. B}\ }\textbf {\bibinfo {volume} {82}},\
  \bibinfo {pages} {045127} (\bibinfo {year} {2010})}\BibitemShut {NoStop}%
\bibitem [{\citenamefont {Anderson}\ \emph {et~al.}(2015)\citenamefont
  {Anderson}, \citenamefont {Wu}, \citenamefont {Boyack},\ and\ \citenamefont
  {Levin}}]{ourTop}%
  \BibitemOpen
  \bibfield  {author} {\bibinfo {author} {\bibfnamefont {B.~M.}\ \bibnamefont
  {Anderson}}, \bibinfo {author} {\bibfnamefont {C.-T.}\ \bibnamefont {Wu}},
  \bibinfo {author} {\bibfnamefont {R.}~\bibnamefont {Boyack}}, \ and\ \bibinfo
  {author} {\bibfnamefont {K.}~\bibnamefont {Levin}},\ }\href {\doibase
  10.1103/PhysRevB.92.134523} {\bibfield  {journal} {\bibinfo  {journal} {Phys.
  Rev. B}\ }\textbf {\bibinfo {volume} {92}},\ \bibinfo {pages} {134523}
  (\bibinfo {year} {2015})}\BibitemShut {NoStop}%
\bibitem [{\citenamefont {Hasan}\ and\ \citenamefont
  {Kane}(2010)}]{RevModPhys.82.3045}%
  \BibitemOpen
  \bibfield  {author} {\bibinfo {author} {\bibfnamefont {M.~Z.}\ \bibnamefont
  {Hasan}}\ and\ \bibinfo {author} {\bibfnamefont {C.~L.}\ \bibnamefont
  {Kane}},\ }\href {\doibase 10.1103/RevModPhys.82.3045} {\bibfield  {journal}
  {\bibinfo  {journal} {Rev. Mod. Phys.}\ }\textbf {\bibinfo {volume} {82}},\
  \bibinfo {pages} {3045} (\bibinfo {year} {2010})}\BibitemShut {NoStop}%
\bibitem [{\citenamefont {Qi}\ and\ \citenamefont
  {Zhang}(2011)}]{RevModPhys.83.1057}%
  \BibitemOpen
  \bibfield  {author} {\bibinfo {author} {\bibfnamefont {X.-L.}\ \bibnamefont
  {Qi}}\ and\ \bibinfo {author} {\bibfnamefont {S.-C.}\ \bibnamefont {Zhang}},\
  }\href {\doibase 10.1103/RevModPhys.83.1057} {\bibfield  {journal} {\bibinfo
  {journal} {Rev. Mod. Phys.}\ }\textbf {\bibinfo {volume} {83}},\ \bibinfo
  {pages} {1057} (\bibinfo {year} {2011})}\BibitemShut {NoStop}%
\bibitem [{\citenamefont {Majorana}(1937)}]{Majorana}%
  \BibitemOpen
  \bibfield  {author} {\bibinfo {author} {\bibfnamefont {E.}~\bibnamefont
  {Majorana}},\ }\href@noop {} {\bibfield  {journal} {\bibinfo  {journal}
  {Nuovo Cimento}\ }\textbf {\bibinfo {volume} {14}},\ \bibinfo {pages} {171}
  (\bibinfo {year} {1937})}\BibitemShut {NoStop}%
\bibitem [{\citenamefont {{Beenakker}}(2011)}]{2011arXiv1112.1950B}%
  \BibitemOpen
  \bibfield  {author} {\bibinfo {author} {\bibfnamefont {C.~W.~J.}\
  \bibnamefont {{Beenakker}}},\ }\href@noop {} {\bibfield  {journal} {\bibinfo
  {journal} {ArXiv e-prints}\ } (\bibinfo {year} {2011})},\ \Eprint
  {http://arxiv.org/abs/1112.1950} {arXiv:1112.1950 [cond-mat.mes-hall]}
  \BibitemShut {NoStop}%
\bibitem [{\citenamefont {Alicea}(2012)}]{0034-4885-75-7-076501}%
  \BibitemOpen
  \bibfield  {author} {\bibinfo {author} {\bibfnamefont {J.}~\bibnamefont
  {Alicea}},\ }\href {http://stacks.iop.org/0034-4885/75/i=7/a=076501}
  {\bibfield  {journal} {\bibinfo  {journal} {Reports on Progress in Physics}\
  }\textbf {\bibinfo {volume} {75}},\ \bibinfo {pages} {076501} (\bibinfo
  {year} {2012})}\BibitemShut {NoStop}%
\bibitem [{\citenamefont {Su}\ \emph {et~al.}(1979)\citenamefont {Su},
  \citenamefont {Schrieffer},\ and\ \citenamefont
  {Heeger}}]{PhysRevLett.42.1698}%
  \BibitemOpen
  \bibfield  {author} {\bibinfo {author} {\bibfnamefont {W.~P.}\ \bibnamefont
  {Su}}, \bibinfo {author} {\bibfnamefont {J.~R.}\ \bibnamefont {Schrieffer}},
  \ and\ \bibinfo {author} {\bibfnamefont {A.~J.}\ \bibnamefont {Heeger}},\
  }\href {\doibase 10.1103/PhysRevLett.42.1698} {\bibfield  {journal} {\bibinfo
   {journal} {Phys. Rev. Lett.}\ }\textbf {\bibinfo {volume} {42}},\ \bibinfo
  {pages} {1698} (\bibinfo {year} {1979})}\BibitemShut {NoStop}%
\bibitem [{\citenamefont {Ebisu}\ \emph {et~al.}(2015)\citenamefont {Ebisu},
  \citenamefont {Yada}, \citenamefont {Kasai},\ and\ \citenamefont
  {Tanaka}}]{PhysRevB.91.054518}%
  \BibitemOpen
  \bibfield  {author} {\bibinfo {author} {\bibfnamefont {H.}~\bibnamefont
  {Ebisu}}, \bibinfo {author} {\bibfnamefont {K.}~\bibnamefont {Yada}},
  \bibinfo {author} {\bibfnamefont {H.}~\bibnamefont {Kasai}}, \ and\ \bibinfo
  {author} {\bibfnamefont {Y.}~\bibnamefont {Tanaka}},\ }\href {\doibase
  10.1103/PhysRevB.91.054518} {\bibfield  {journal} {\bibinfo  {journal} {Phys.
  Rev. B}\ }\textbf {\bibinfo {volume} {91}},\ \bibinfo {pages} {054518}
  (\bibinfo {year} {2015})}\BibitemShut {NoStop}%
\bibitem [{\citenamefont {Tinkham}(1996)}]{tinkham}%
  \BibitemOpen
  \bibfield  {author} {\bibinfo {author} {\bibfnamefont {M.}~\bibnamefont
  {Tinkham}},\ }\href@noop {} {\emph {\bibinfo {title} {Introduction to
  Superconductivity}}}\ (\bibinfo  {publisher} {Dover Publications, Inc},\
  \bibinfo {address} {Mineola, New York},\ \bibinfo {year} {1996})\BibitemShut
  {NoStop}%
\bibitem [{sup()}]{supp}%
  \BibitemOpen
  \href@noop {} {}\bibinfo {note} {For details, see the Supplemental material
  which includes discussion of the numerical algorithm, finite size effects and
  the analytics underlying the Su-Schrieffer-Heeger
  correspondence.}\BibitemShut {Stop}%
\bibitem [{\citenamefont {Sato}\ \emph {et~al.}(2011)\citenamefont {Sato},
  \citenamefont {Tanaka}, \citenamefont {Yada},\ and\ \citenamefont
  {Yokoyama}}]{PhysRevB.83.224511}%
  \BibitemOpen
  \bibfield  {author} {\bibinfo {author} {\bibfnamefont {M.}~\bibnamefont
  {Sato}}, \bibinfo {author} {\bibfnamefont {Y.}~\bibnamefont {Tanaka}},
  \bibinfo {author} {\bibfnamefont {K.}~\bibnamefont {Yada}}, \ and\ \bibinfo
  {author} {\bibfnamefont {T.}~\bibnamefont {Yokoyama}},\ }\href {\doibase
  10.1103/PhysRevB.83.224511} {\bibfield  {journal} {\bibinfo  {journal} {Phys.
  Rev. B}\ }\textbf {\bibinfo {volume} {83}},\ \bibinfo {pages} {224511}
  (\bibinfo {year} {2011})}\BibitemShut {NoStop}%
\bibitem [{\citenamefont {Black-Schaffer}\ and\ \citenamefont
  {Balatsky}(2012)}]{PhysRevB.86.144506}%
  \BibitemOpen
  \bibfield  {author} {\bibinfo {author} {\bibfnamefont {A.~M.}\ \bibnamefont
  {Black-Schaffer}}\ and\ \bibinfo {author} {\bibfnamefont {A.~V.}\
  \bibnamefont {Balatsky}},\ }\href {\doibase 10.1103/PhysRevB.86.144506}
  {\bibfield  {journal} {\bibinfo  {journal} {Phys. Rev. B}\ }\textbf {\bibinfo
  {volume} {86}},\ \bibinfo {pages} {144506} (\bibinfo {year}
  {2012})}\BibitemShut {NoStop}%
\bibitem [{\citenamefont {Asano}\ and\ \citenamefont
  {Tanaka}(2013)}]{PhysRevB.87.104513}%
  \BibitemOpen
  \bibfield  {author} {\bibinfo {author} {\bibfnamefont {Y.}~\bibnamefont
  {Asano}}\ and\ \bibinfo {author} {\bibfnamefont {Y.}~\bibnamefont {Tanaka}},\
  }\href {\doibase 10.1103/PhysRevB.87.104513} {\bibfield  {journal} {\bibinfo
  {journal} {Phys. Rev. B}\ }\textbf {\bibinfo {volume} {87}},\ \bibinfo
  {pages} {104513} (\bibinfo {year} {2013})}\BibitemShut {NoStop}%
\bibitem [{\citenamefont {Stanev}\ and\ \citenamefont
  {Galitski}(2014)}]{Galitskinew}%
  \BibitemOpen
  \bibfield  {author} {\bibinfo {author} {\bibfnamefont {V.}~\bibnamefont
  {Stanev}}\ and\ \bibinfo {author} {\bibfnamefont {V.}~\bibnamefont
  {Galitski}},\ }\href {\doibase 10.1103/PhysRevB.89.174521} {\bibfield
  {journal} {\bibinfo  {journal} {Phys. Rev. B}\ }\textbf {\bibinfo {volume}
  {89}},\ \bibinfo {pages} {174521} (\bibinfo {year} {2014})}\BibitemShut
  {NoStop}%
\bibitem [{\citenamefont {Albrecht}\ \emph {et~al.}(2016)\citenamefont
  {Albrecht}, \citenamefont {Higginbotham}, \citenamefont {Madsen},
  \citenamefont {Kuenmmeth}, \citenamefont {Jespersen}, \citenamefont {Nygard},
  \citenamefont {Krogstrup},\ and\ \citenamefont {Marcus}}]{Marcus}%
  \BibitemOpen
  \bibfield  {author} {\bibinfo {author} {\bibfnamefont {S.~M.}\ \bibnamefont
  {Albrecht}}, \bibinfo {author} {\bibfnamefont {A.~P.}\ \bibnamefont
  {Higginbotham}}, \bibinfo {author} {\bibfnamefont {M.}~\bibnamefont
  {Madsen}}, \bibinfo {author} {\bibfnamefont {F.}~\bibnamefont {Kuenmmeth}},
  \bibinfo {author} {\bibfnamefont {T.~S.}\ \bibnamefont {Jespersen}}, \bibinfo
  {author} {\bibfnamefont {J.}~\bibnamefont {Nygard}}, \bibinfo {author}
  {\bibfnamefont {P.}~\bibnamefont {Krogstrup}}, \ and\ \bibinfo {author}
  {\bibfnamefont {C.~M.}\ \bibnamefont {Marcus}},\ }\href {\doibase
  10.1038/nature17162} {\bibfield  {journal} {\bibinfo  {journal} {Nature}\
  }\textbf {\bibinfo {volume} {531}},\ \bibinfo {pages} {206} (\bibinfo {year}
  {2016})}\BibitemShut {NoStop}%
\bibitem [{\citenamefont {Schnyder}\ and\ \citenamefont
  {Brydon}(2015)}]{Schnyder}%
  \BibitemOpen
  \bibfield  {author} {\bibinfo {author} {\bibfnamefont {A.~P.}\ \bibnamefont
  {Schnyder}}\ and\ \bibinfo {author} {\bibfnamefont {P.~M.~R.}\ \bibnamefont
  {Brydon}},\ }\href {http://stacks.iop.org/0953-8984/27/i=24/a=243201}
  {\bibfield  {journal} {\bibinfo  {journal} {Journal of Physics: Condensed
  Matter}\ }\textbf {\bibinfo {volume} {27}},\ \bibinfo {pages} {243201}
  (\bibinfo {year} {2015})}\BibitemShut {NoStop}%
\bibitem [{\citenamefont {Nesterov}\ \emph {et~al.}(2016)\citenamefont
  {Nesterov}, \citenamefont {Houzet},\ and\ \citenamefont
  {Meyer}}]{AnomJosephson}%
  \BibitemOpen
  \bibfield  {author} {\bibinfo {author} {\bibfnamefont {K.~N.}\ \bibnamefont
  {Nesterov}}, \bibinfo {author} {\bibfnamefont {M.}~\bibnamefont {Houzet}}, \
  and\ \bibinfo {author} {\bibfnamefont {J.~S.}\ \bibnamefont {Meyer}},\ }\href
  {\doibase 10.1103/PhysRevB.93.174502} {\bibfield  {journal} {\bibinfo
  {journal} {Phys. Rev. B}\ }\textbf {\bibinfo {volume} {93}},\ \bibinfo
  {pages} {174502} (\bibinfo {year} {2016})}\BibitemShut {NoStop}%
\end{thebibliography}%

\end{document}